\documentclass{aa}  
\usepackage{graphicx}
\usepackage{txfonts}
\usepackage{xcolor} 
\usepackage{hhline}
\usepackage{longtable}
\usepackage[colorlinks=true, linkcolor=blue, urlcolor=blue,citecolor=blue, anchorcolor=blue]{hyperref}
\usepackage{svg}
\begin{document}

   \title{Dynamical masses of two young transiting sub-Neptunes orbiting HD\,63433}

   \author{M.\ Mallorqu\'in \inst{1,2} 
          \and
          V.\ J.\ S.\ B\'ejar \inst{1,2} 
          \and
          N.\ Lodieu  \inst{1,2} 
          \and
          M.\ R.\ Zapatero Osorio \inst{3} 
          \and
          H.\ Tabernero \inst{3} 
          \and
           A.\ Su\'arez Mascare\~no \inst{1,2} 
          \and
          M.\ Zechmeister \inst{4} 
          \and
          R.\ Luque \inst{5} 
          \and
          E.\ Pall\'e \inst{1,2} 
          \and          
          D.\ Montes \inst{6} 
          }



   \institute{Instituto de Astrof\'isica de Canarias (IAC), Calle V\'ia L\'actea s/n, E-38205 La Laguna, Tenerife, Spain
             \email{mmd@iac.es}
         \and
         Departamento de Astrof\'isica, Universidad de La Laguna (ULL), E-38206 La Laguna, Tenerife, Spain
         \and
         Centro de Astrobiolog\'ia (CSIC-INTA), Ctra. Ajalvir km 4, 28850, Torrej\'on de Ardoz, Madrid, Spain
         \and
         Institut f\"ur Astrophysik, Georg-August-Universit\"at, Friedrich-Hund-Platz 1, 37077 G\"ottingen, Germany
         \and 
         Department of Astronomy \& Astrophysics, University of Chicago, Chicago, IL 60637, USA.
         \and
         Departamento de F\'isica de la Tierra y Astrof\'isica \& IPARCOS-UCM (Instituto de F\'isica de Part\'iculas y del Cosmos de la UCM),
Facultad de Ciencias F\'isicas, Universidad Complutense de Madrid, 28040 Madrid, Spain
             }

   \date{Received \today{}; accepted \today{}}

 
  \abstract
   {
   Although the number of exoplanets reported in the literature exceeds 5000 so far, only a few dozen of them are young planets ($\le$900 Myr). However, a complete characterization of these young planets is key to understanding the current properties of the entire population. Hence, it is necessary to constrain the planetary formation processes and the timescales of dynamical evolution by measuring the masses of exoplanets transiting young stars.
   }
   {
   We characterize and measure the masses of two transiting planets orbiting the 400 Myr old solar-type star HD\,63433, which is a member of the Ursa Major moving group.
   }
   {
   We analysed precise photometric light curves of five sectors of the TESS mission with a baseline of $\sim$750 days and obtained $\sim$150 precise radial velocity measurements with the visible and infrared arms of the CARMENES instrument at the Calar Alto 3.5 m telescope in two different campaigns of $\sim$500 days. We performed a combined photometric and spectroscopic analysis to retrieve the planetary properties of two young planets. The strong stellar activity signal was modelled by Gaussian regression processes.
   }
   {
   We have updated the transit parameters of HD\,63433\,b and c and obtained planet radii of R$_p^b$\,=\,2.140\,$\pm$\,0.087 R$_\oplus$ and R$_p^c$\,=\,2.692\,$\pm$\,0.108 R$_\oplus$. Our analysis allowed us to determine the dynamical mass of the outer planet with a 4$\sigma$ significance ($M_p^c$\,=\,15.54\,$\pm$\,3.86 M$_\oplus$) and set an upper limit on the mass of the inner planet at 3$\sigma$ ($M_p^b$\,$<$\,21.76 M$_\oplus$). According to theoretical models, both planets are expected to be sub-Neptunes, whose interiors mostly consist of silicates and water with no dominant composition of iron, and whose gas envelopes are lower than 2\% in the case of HD\,63433\,c. The envelope is unconstrained in HD\,63433\,b.
   }
   {}


   \maketitle
%

\section{Introduction}
\label{sec:intro}

Despite the discovery of thousands of exoplanets, the mechanisms of planet formation and evolution have not been well tested with observations, mainly because of the lack of a complete photometric and spectroscopic characterization of young planets. The determination of their physical parameters is a great challenge because the stellar activity of host young stars is high. The activity is produced by a rapid rotation and strong magnetic activity, which, in most cases, produce photometric and radial velocity variations that are several times greater than the signals assigned to the planets. This also hampers the search for new young planet candidates.

One of the critical questions of the last years in exoplanet research is the existence of a bimodal distribution in the size of the small planets, that is,\ the so-called radius valley or Fulton gap \citep{fulton17}. Several evolution models have been proposed to explain this gap by considering that a single population of rocky planets (with a similar composition to that of Earth) is formed with a gaseous envelope that is sculpted by evolution through rapid loss of its atmosphere ($\text{about several tens}$ of million years) by photoevaporation \citep{owen17} or slowly ($\sim$1 Gyr) by core-powered mass loss \citep{ginzburg18}. However, the recent study by \cite{luque22} shows that in M dwarfs, the planets around the valley are distributed based on their composition (i.e.\ densities) into three classes: rocky, water-rich (i.e.\ planets that are 50\% rocky and 50\% water ice by mass), and gas-rich planets (i.e.\ could be rocky planets with massive H/He envelopes or water-rich planets with less massive envelopes). According to the models of \citet{venturini20}, the formation of water and gaseous planets occurs beyond the ice line, and the planets later migrate inwards. The radius valley is explained in these models by the initial differences in the core sizes below or beyond ice lines and the subsequent loss of the envelopes of gas planets that migrated outside-in. To constrain the planet formation and evolution models, it is therefore critical to measure the density of planets with sizes between those of Earth and Neptune at different ages below 1 Gyr to determine whether most of the super-Earths are rocky planets from their origin, or if they are born as small gaseous planets that lose their atmospheres.

Recently, several planets with sizes between those of Earth and Neptune have been found orbiting stellar members of the Hyades \citep{mann16, ciar18, mann18}, the Praesepe \citep{obermeier16, mann17, rizzuto18} and $\delta$\,Lyr \citep{bouma22a} open clusters, in the Ursa Major \citep{mann20}, $\beta$\,Pic \citep{plav20}, Pisces-Eridanus \citep{newton21}, Melange-1 \citep{toff21}, and AB Doradus \citep{zhou22} young moving groups, in the Cep-Her complex \citep{bouma22b}, and around other young field stars \citep{david18a, david18b, sun19, zhou21, ment21, hedges21, kossak21, barr22, vach22} during the K2 and TESS missions. Only six of these young planetary systems have measured masses and accordingly, densities, derived from dedicated radial velocity (RV) campaigns: AU\,Mic\,b and c ($\sim$20 Myr; \citet{klein21, cale21, zicher22, klein22}), TOI-1807\,b ($\sim$300 Myr; \citet{nardiello22}), TOI-560\,b and c ($\sim$490 Myr; \citet{barr22, elmufti22}), K2-25\,b ($\sim$725 Myr; \citet{stef20}), K2-100\,b ($\sim$750 Myr; \citet{barr19}), and TOI-1201\,b (600--800 Myr; \citet{kossak21}). 

HD\,63433 (TOI-1726, V377 Gem) is a bright ($V$\,=\,6.9 mag) solar-type star member of the Ursa Major moving group \citep{montes2001b} with an estimated age of 414\,$\pm$\,23 Myr \citep{jones15}. Its two transiting exoplanets have orbital periods and radii of 7.11 days and 2.15\,$\pm$\,0.10 R$_\oplus$ for the inner planet, and 20.55 days and 2.67\,$\pm$\,0.12 R$_\oplus$ for the outer planet \citep{mann20}. Furthermore, by modelling the Rossiter-McLaughlin (RM) effect \citep{rossiter, mclaughlin}, HD\,63433\,b and c have been measured a prograde orbits with a sky-projected obliquity of $\lambda$\,=\,1.0$_{-43.0}^{+41.0\circ}$ \citep{mann20} and $\lambda$\,=\,$-$1.0$_{-32.0}^{+35.0\circ}$ \citep{fei20}, respectively. The atmospheric escape has also been studied for both planets, with the detection of Ly\,$\alpha$ absorption in HD\,63433\,c but not in HD\,63433\,b, whereas no helium absorption (10833 $\dot{A}$) was detected in either planet \citep{zhang22}. These results seem to indicate that the two planets have different atmospheric compositions \citep{zhang22}.

In this paper, we present the mass determination of HD\,63433\,c and an upper limit for the mass of HD\,63433\,b. The paper is organised as follows. In Section \ref{sec:obs} we describe the photometry of the Transiting Exoplanet Survey Satellite (TESS) and ground based follow-up observations of the system. In Section \ref{sec:properties} we revise the physical properties of the star. In Sections \ref{sec:tr} and \ref{sec:rv}, we perform a transit and RV analysis of the planets, respectively. A joint-fit modelling of the photometric and RV time-series is carried out in Section \ref{sec:joint}. In Section \ref{sec:disc} we discuss the composition of the planet and the main implications for theoretical models of exoplanets. We summarize our main results in Section \ref{sec:concl}. 

\section{Observations}
\label{sec:obs}

\subsection{TESS photometry}
\label{sec:tess}

HD\,63433 was observed by TESS in five sectors with a 2-minute short cadence. The sector used for the discovery \citep{mann20}, sector 20, ran from 24 December 2019 until 21 January 2020 as part of cycle 2 of the TESS primary mission. The other four sectors (44 to 47) were contiguous and provided a light curve between 12 October 2021 and 28 January 2022 during cycle 4 of the first extended mission. At the time of this publication, the target is not expected to be observed again with TESS\@. 

\begin{figure*}[ht!]
\begin{center}
\includegraphics[width=0.6\linewidth]{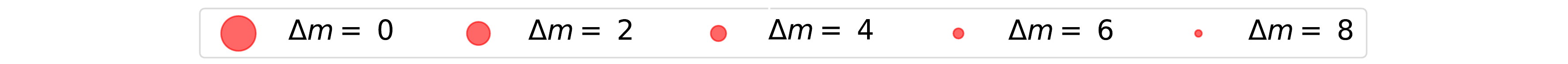}\\
\includegraphics[width=0.17\linewidth]{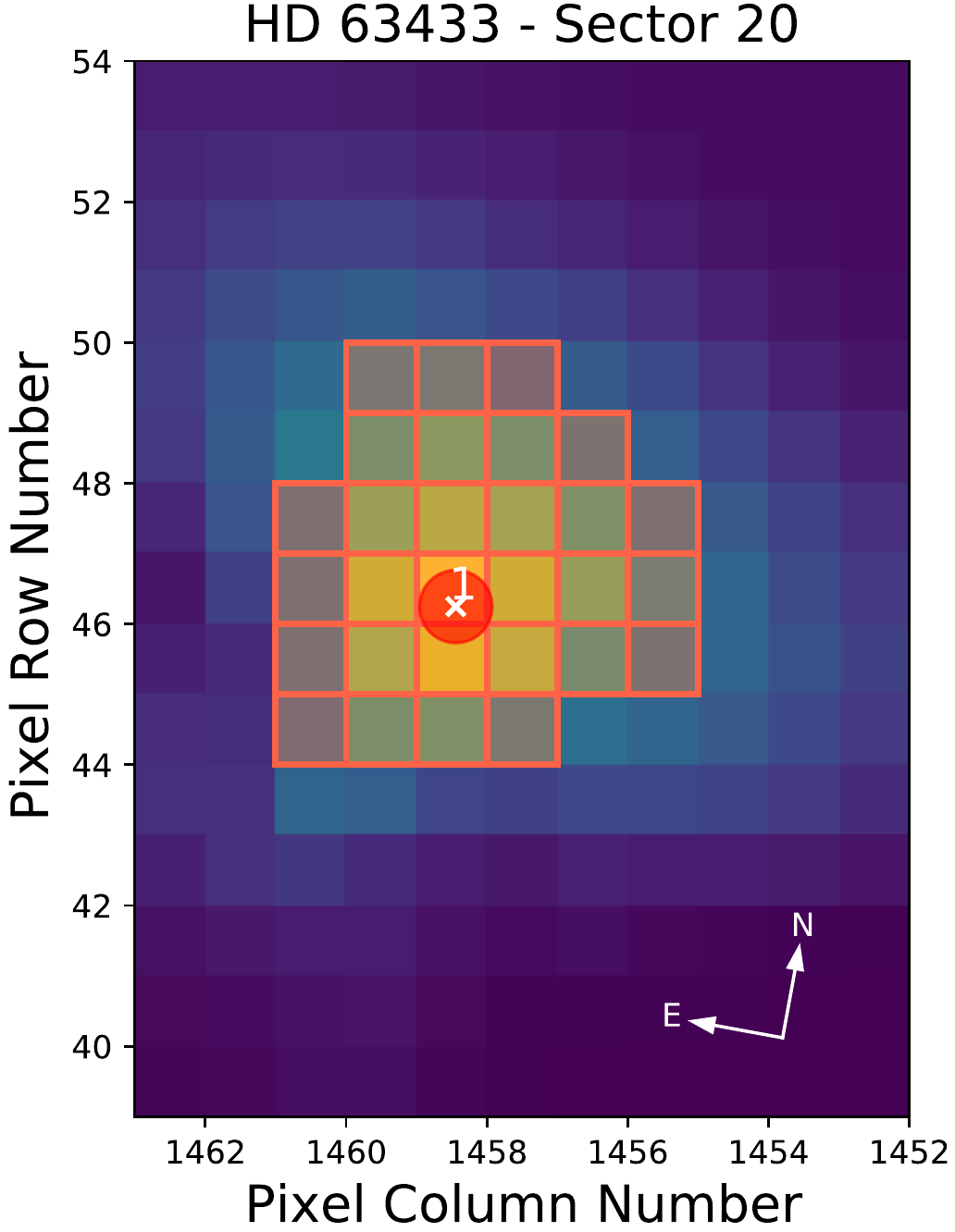}
\includegraphics[width=0.175\linewidth]{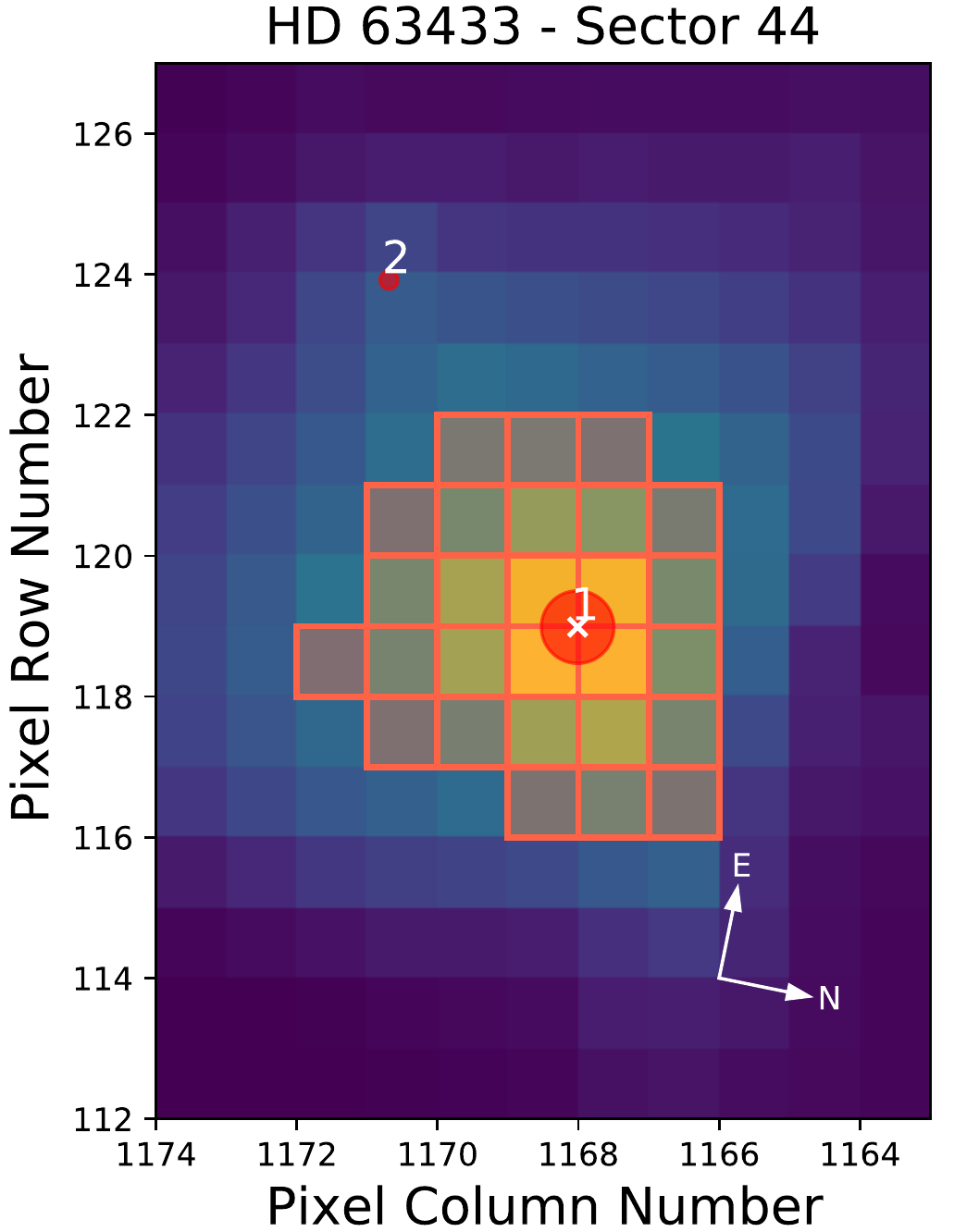}
\includegraphics[width=0.18\linewidth]{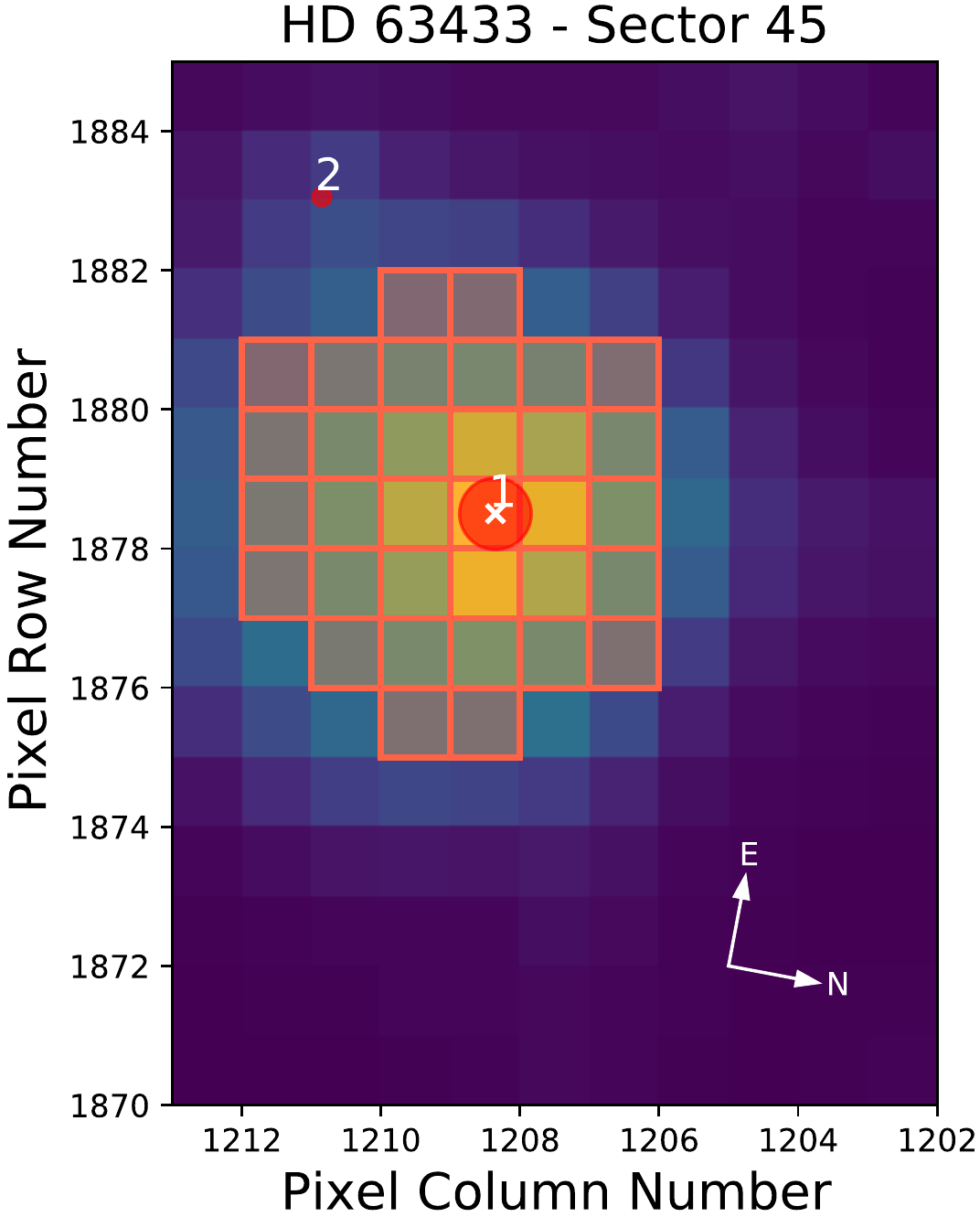}
\includegraphics[width=0.18\linewidth]{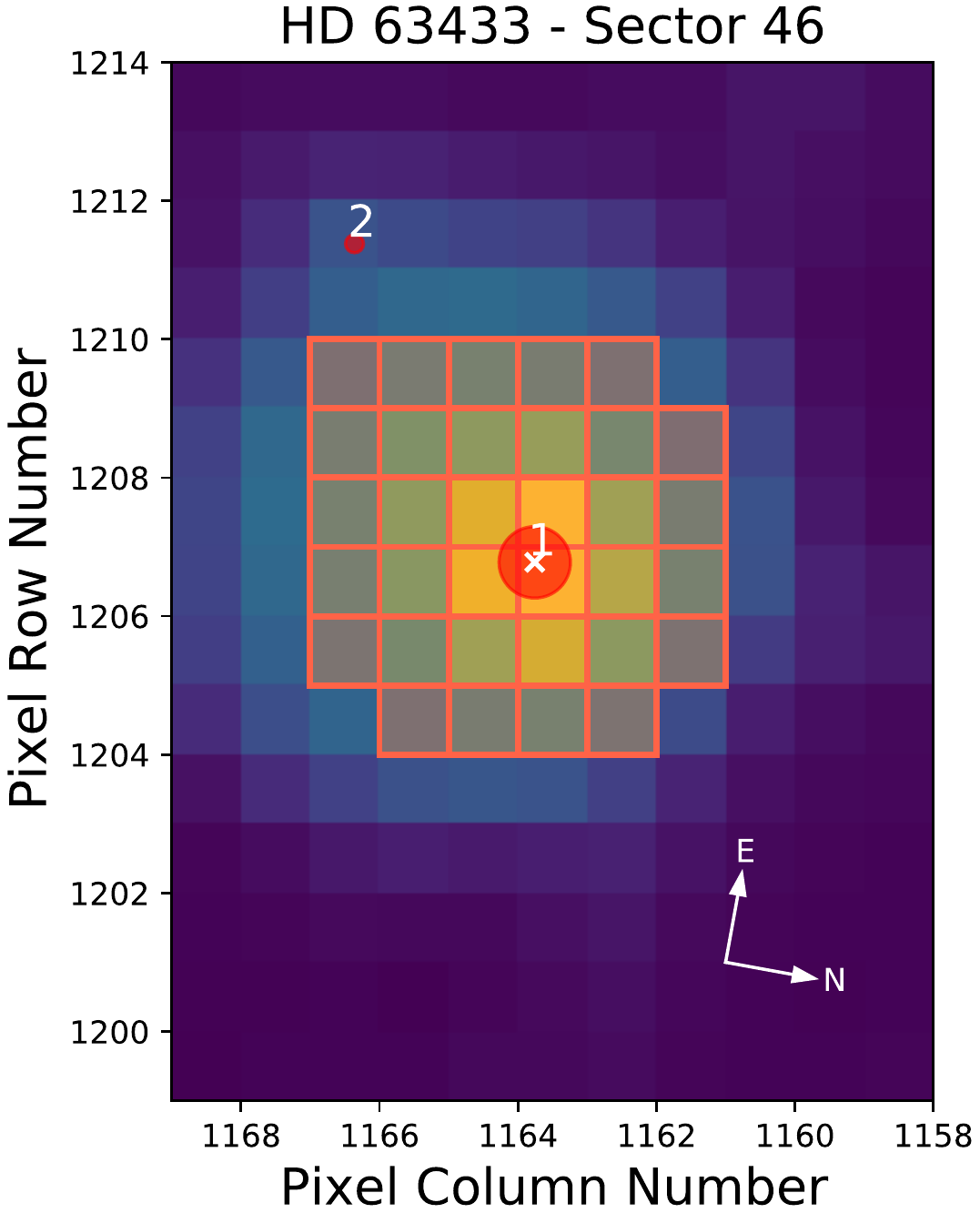}
\includegraphics[width=0.24\linewidth]{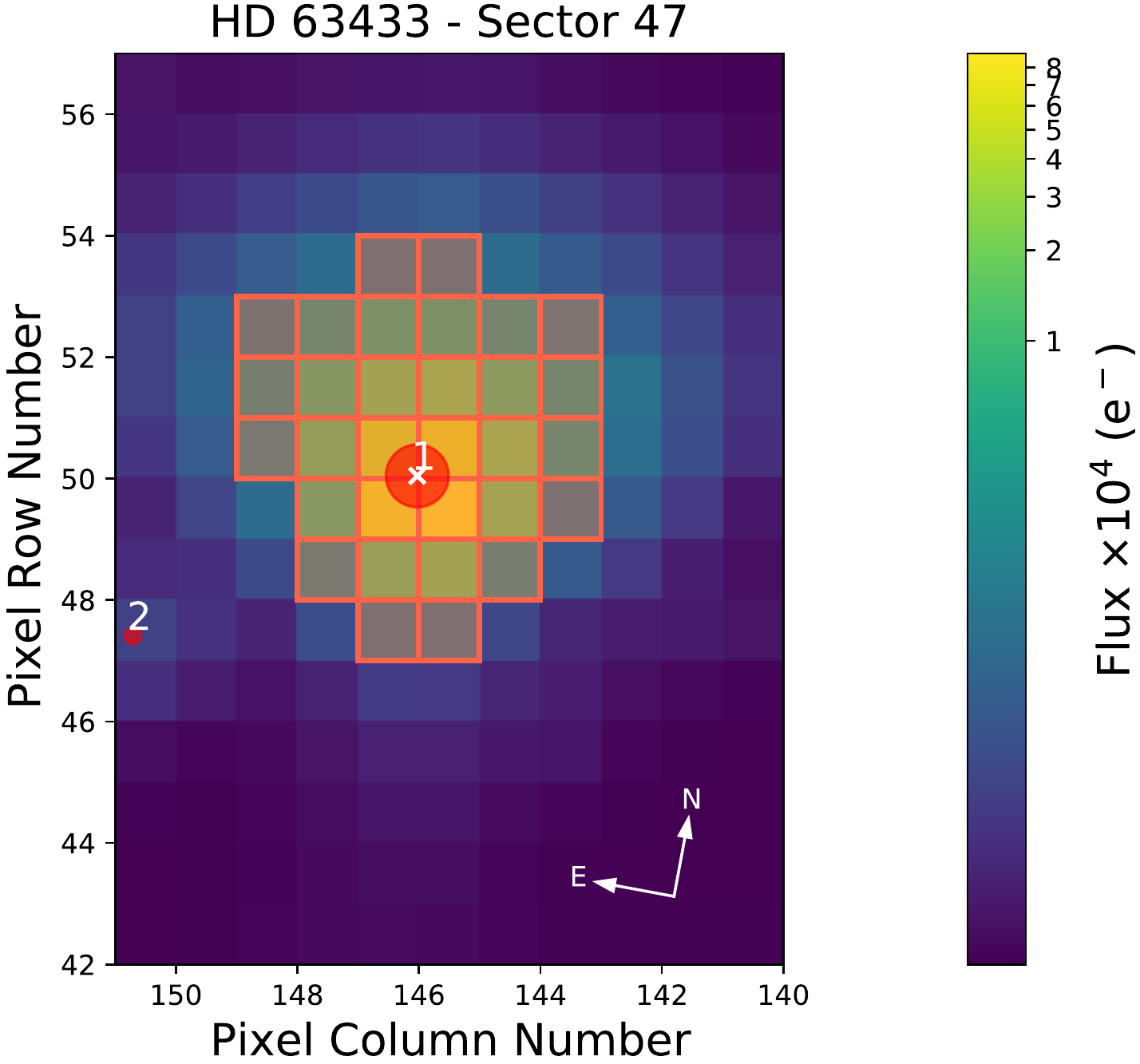}

\caption{TESS TPF plots for HD\,63433 (white crosses). The red squares indicate the best optimal photometric aperture used to obtain the SAP flux. G-band magnitudes from $Gaia$ DR3 are shown with different sizes of red circles for all nearby stars with respect to HD\,63433 up to 8 magnitudes fainter.
\label{fig:TPF}}
\end{center}
\end{figure*}

\begin{figure*}[ht!]
\includegraphics[width=1\linewidth]{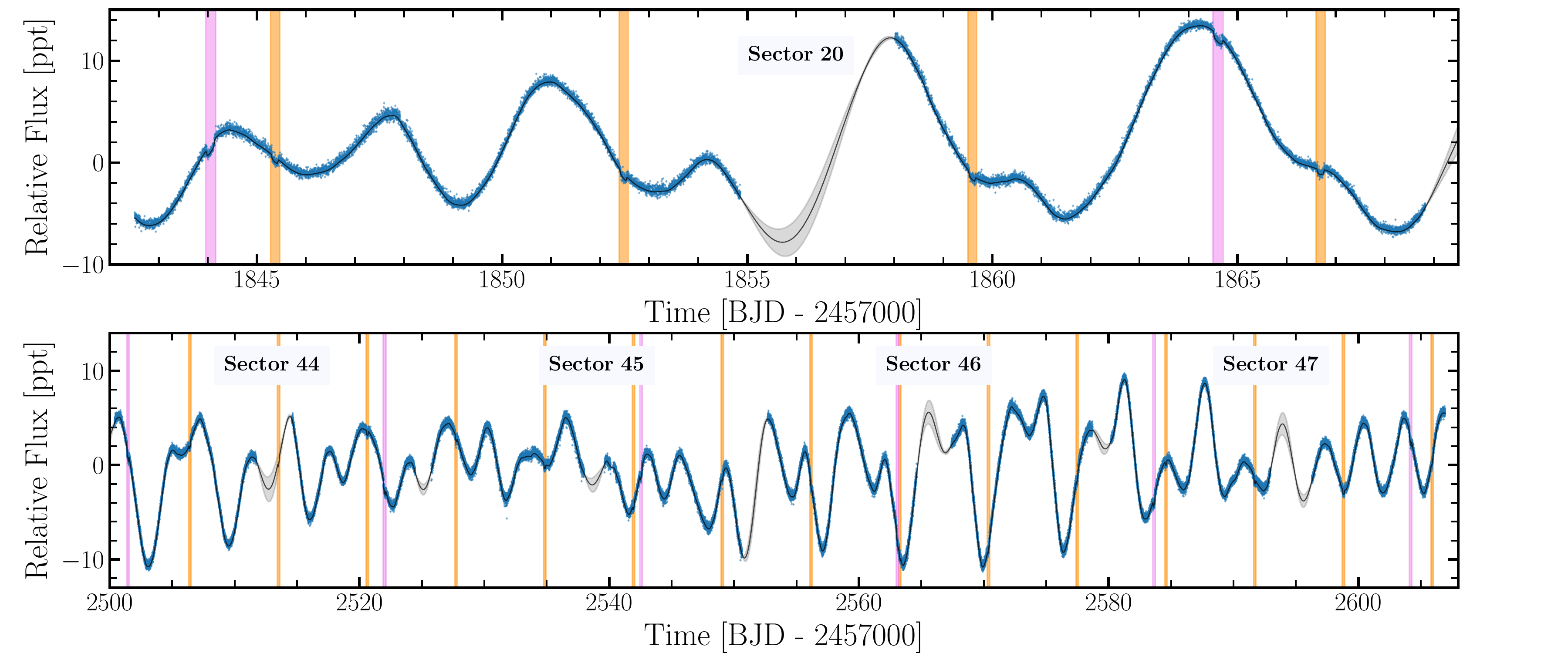}
\caption{Light curves of HD\,63433 for the five TESS sectors. The blue dots correspond to the PDCSAP flux data. The black line and the grey shaded region indicate the best model and its 1$\sigma$ uncertainty, respectively. The vertical lines show the times of the planetary transits for HD\,63433\,b (orange) and HD\,63433\,c (purple).
\label{fig:LC_TESS}}
\end{figure*}

All sectors were processed by the Science Processing Operations Center (SPOC; \citealp{SPOC}) photometry and transit search pipeline at the NASA Ames Research Center. The light curves and target pixel files (TPFs) were downloaded from the Mikulski Archive for Space Telescopes\footnote{\url{https://archive.stsci.edu/}} (MAST), which  provides the simple aperture photometry (SAP) and the pre-search data conditioning SAP flux (PDCSAP). To confirm that the best photometric aperture for the TESS light curve does not include any additional bright sources down to 8 mag fainter and contaminates the light curve and the transit depth, we plot in Fig.\,\ref{fig:TPF} the TPF using \texttt{tpfplotter}\footnote{\url{https://github.com/jlillo/tpfplotter}} \citep{aller2020}, which  produces the best SAP flux. We also overplot the $Gaia$ Data Release 3 (DR3) catalogue \citep{gaiadr3} on top of the TPF. We searched for possible sources of contamination and verified that no source lay close to the chosen photometric aperture. For the rest of our analysis, we  used the PDCSAP flux, which was corrected for instrumental errors and crowding. However, we tested that our analysis and results using SAP flux light curves are similar. The PDCSAP flux light curve for all sectors of HD\,63433 is illustrated in Fig.\,\ref{fig:LC_TESS} along with the best-fit model (see Sect.\ \ref{sec:joint} for details). The light curve has a dispersion of $\sigma_{\mathrm{TESS}}$\,$\sim$\,4.3 parts per thousand (ppt), an average error bar of $\sim$0.2 ppt, and a modulation with peak-to-peak variations up to 20 ppt whose amplitude changes significantly from one period to the next. Moreover, the light curve presents a few flares with amplitudes smaller than 1.5 ppt that were removed with the following procedure: first, we created a smoothed model using a Savitzky–Golay filter. After removing the model from the light curve, we iteratively removed all values higher than 3 times the root mean square of the residuals. We verified that most of the flares were removed through this process.
 
\subsection{LCO photometry}
\label{sec:lco}

To monitor the photometric stellar activity contemporaneously with the RV, we observed HD\,63433 using the 40\,cm telescopes of Las Cumbres Observatory Global Telescope \citep[\textit{LCOGT};][]{Brown2013} in the $V$--band at the Teide, McDonald, and Haleakala observatories between 15 September 2020 and 28 December 2020, and between 14 September 2021 and 30 March 2022. We obtained 98 and 176 observing epochs, 82 and 141 of which were good, respectively, and we acquired 30 and 50 individual exposures of 1/2 s per epoch in the 40\,cm telescopes. The 40\,cm telescopes are equipped with a 3k$\times$2k SBIG CCD camera with a pixel scale of 0.571\,arcsec, providing a field of view of 29.2$\times$19.5\,arcmin. Weather conditions at observatories were mostly clear, and the average seeing varies from 1.0\arcsec to 3.0\arcsec. Raw data were processed using the BANZAI pipeline \citep{McCully2018}, which includes bad pixel, bias, dark, and flat-field corrections for each individual night. We performed aperture photometry in the $V$--band for HD\,63433 and two reference stars of the field and obtained the relative differential photometry between the target and the brightest reference. We adopted an aperture of 10 pixels ($\sim$6\arcsec), which minimizes the dispersion of the differential light curve. The light curve has a dispersion of $\sigma_{\mathrm{LCO}}$\,$\sim$\,46 mmag, and the mean value of the uncertainties is $\sim$17 mmag.

\subsection{CARMENES}
\label{sec:carm}

Between 19 September 2020 and 23 February 2022, we collected 157 high-resolution spectra, divided into two campaigns, with the Calar Alto high-Resolution search for M dwarfs with Exoearths with Near-infrared and optical Echelle Spectrographs (CARMENES) mounted on the 3.5\,m telescope at Calar Alto Observatory, Almer\'ia (Spain), under the observing programs H20-3.5-027, 21B-3.5-015, and 22A-3.5-009. To properly model the stellar activity, we designed an observational strategy to obtain three to five spectra per stellar rotation period ($\sim$6.4 days; see Sect.\,\ref{sec:prot} for details). Finally, we obtained about four spectra per stellar rotation on average.

\begin{figure*}[ht!]
\includegraphics[width=1\linewidth]{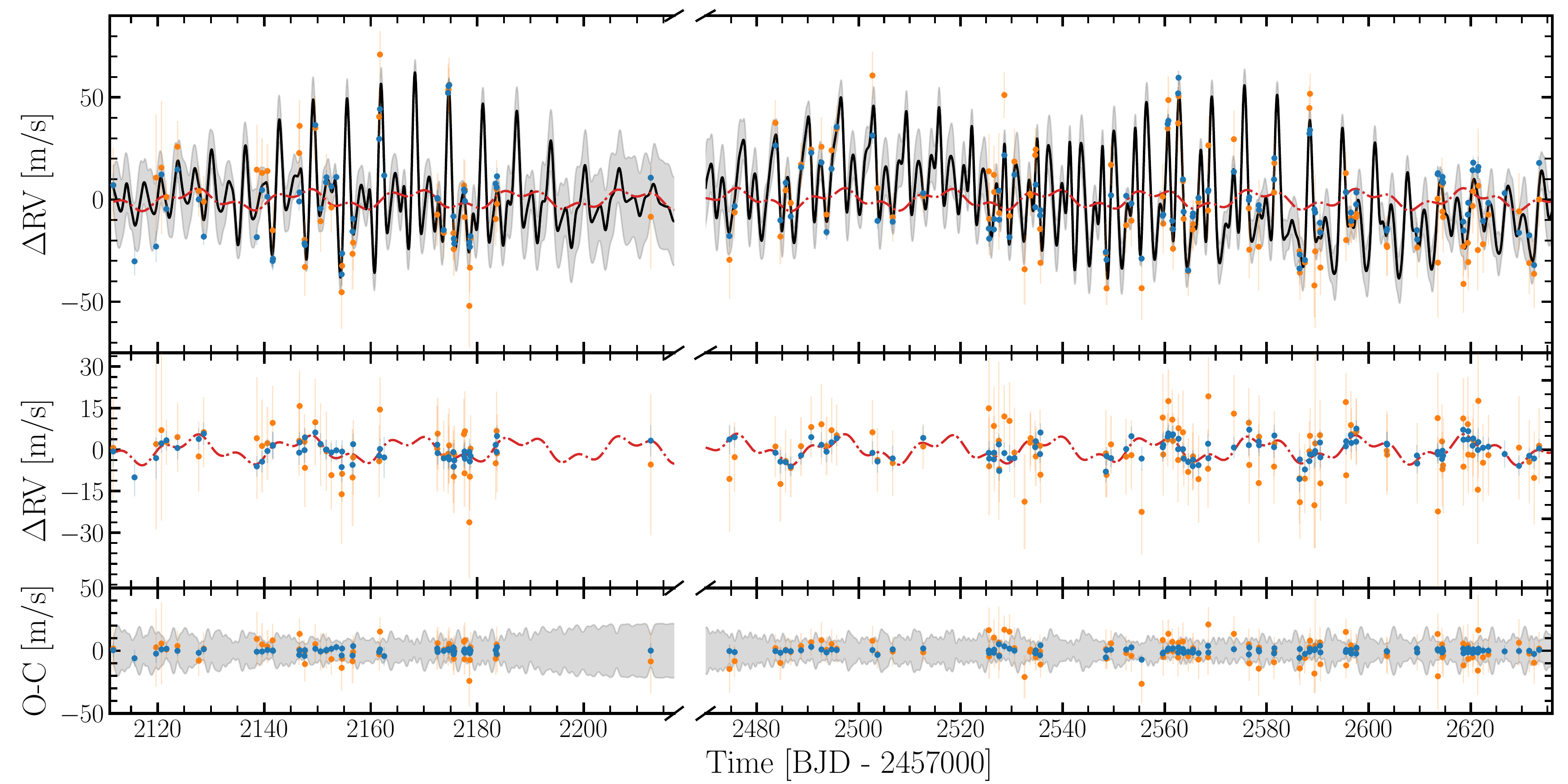}
\caption{CARMENES RV data for HD\,63433 (blue and orange dots for the visible and infrared, respectively). \textit{Top panel}: Combined model (black line) with its 1$\sigma$ level of confidence (grey shadow), and the dashed red line depicts the Keplerian model for two planets. \textit{Middle panel}: Keplerian model alone (dashed red line) and CARMENES VIS and NIR data after subtracting the best activity model. \textit{Bottom panel}: Residuals for the best-fit.  
\label{fig:RV_CARM}}
\end{figure*}

The CARMENES spectrograph has two channels \citep{CARMENES, CARMENES18}. The visible (VIS) channel covers the spectral range 520--960 nm, and the near-infrared (NIR) channel covers the spectral range 960--1710 nm with a spectral resolution of $\mathcal{R}$\,=\,94000 and $\mathcal{R}$\,=\,80400. One spectrum of each arm was ruled out because the drift correction was missing. Moreover, another six spectra were discarded in each arm due to their low signal-to-noise ratio (S/N\,$<$\,50). The remaining observations were taken with exposure times of 150 s, resulting in S/Ns per pixel in the range of 68--250 at 745 nm. We used the VIS and NIR channel observations to derive RV measurements. The CARMENES performance, data reduction, and wavelength calibration were made using CARACAL and are described in \citet{2018A&A...609A.117T} and \citet{2018A&A...618A.115K}. Relative RV values and activity indexes such as the H$\alpha$ index, the \ion{Ca}{II} IR triplet (IRT), and the \ion{Na}{I}\,D values were obtained using the {\tt serval}\footnote{\url{https://github.com/mzechmeister/serval}} pipeline \citep{2018A&A...609A..12Z}. Furthermore, the CRX and dLW activity indicators were also computed, where the first indicator takes into account the dependence of the RV on the wavelength, and the second evaluates the variations in the widths of the lines with respect to the reference lines. In addition, this software also produces a high S/N template spectrum by co-adding all the observed spectra after correcting them for the wavelength shifts. The RV measurements were corrected for barycentric motion, secular acceleration, instrumental drift, and nightly zero-points. We searched for possible outliers in the datasets and discarded 4 points in the NIR time-series. This resulted in 150 and 146 RV data points in the VIS and NIR channels, respectively. The typical dispersion of the RV measurements is $\sigma_{\mathrm{CARMENES\ VIS}}$\,$\sim$\,19.8 m\,s$^{-1}$ for the VIS range and $\sigma_{\mathrm{CARMENES\ NIR}}$\,$\sim$\,25.0 m\,s$^{-1}$ in the NIR range. The uncertainties are in the range of 2.6--7.5 m\,s$^{-1}$ with a mean value of 3.9\,m\,s$^{-1}$ in the VIS arm and in the range of 8--38\,m\,s$^{-1}$ with a mean value of 14\,m\,s$^{-1}$ in the NIR. The RV curve for both datasets is illustrated in Fig.\,\ref{fig:RV_CARM} with its best-fit model (see Sect.\ \ref{sec:joint} for details). We searched for RV measurements affected by flares by measuring the relative intensity of a set of spectral lines that are usually associated with stellar activity (H$\alpha$, \ion{Ca}{II}\,IRT, \ion{Na}{I}, and \ion{K}{I}). We compared them with each other to look for significant variations, but none of the spectra lines seem to be affected by flares. In addition, because a significant part of our RV data is contemporaneous with the TESS photometry, we visually inspected whether any of the flares found in the TESS light curve coincided with the RV data. This was not the case, therefore we kept all RV data. In the appendices, Tables \ref{tab:RV_CARMV} and \ref{tab:RV_CARMN} give the time stamps of the spectra in BJD$_{\mathrm{TDB}}$ and the relative RVs measured with {\tt serval} along with their $1\sigma$ error bars.

\section{Stellar properties}
\label{sec:properties}

\subsection{Physical parameters of HD\,63433}

HD\,63433 is a G5V star \citep{gray03} located at a distance of 22.34\,$\pm$\,0.04 pc. The distance was determined from the $Gaia$ DR3 parallax \citep{gaiadr3}. HD\,63433 was first identified as a member of the Ursa Major moving group by \citet{gaidos98} and was later confirmed by \citet{gagne18}. By measuring the lithium abundance and rotation period of HD\,63433, \citet{mann20} confirmed the young age of the star and the membership to the group. We adopt  the recent age determination of 414\,$\pm$\,23 Myr for the
HD\,63433 system that was provided by \citet{jones15} for the Ursa Major moving group.

Using the stellar template, which combines all the 150 spectra, we computed the stellar atmospheric parameters of HD\,63433 with the {\sc SteParSyn} code\footnote{\url{https://github.com/hmtabernero/SteParSyn/}} \citep{tab22}. This code implements the spectral synthesis method with the {\tt emcee}\footnote{\url{https://github.com/dfm/emcee}} Python package \citep{emcee} to retrieve the stellar atmospheric parameters. We employed a grid of synthetic spectra computed with the Turbospectrum \citep{ple12} code, the MARCS stellar atmospheric models \citep{gus08}, and the atomic and molecular data of the $Gaia$-ESO line list \citep{hei21}. We employed a set of \ion{Fe}{i,ii} features that are well suited for analysing FGKM stars \citep{tab22}. We retrieved the following parameters: $T_{\rm eff}$\,$=$\,5553\,$\pm$\,56\,K, $\log{g}$\,$=$\,4.56\,$\pm$\,0.08 dex, [Fe/H]\,$=$\,$-$0.07\,$\pm$\,0.03 dex, and $\varv \sin{i}$\,$=$\,7.76\,$\pm$\,0.10 km~s$^{-1}$.

We estimated the luminosity of HD\,63433 by integrating the observed fluxes from the UV-optical to mid-infrared using VOSA \citep{bayo08}, including the Galaxy Evolution Explorer (GALEX; \citet{bianchi2017}), Tycho \citep{hog2000}, Gaia \citep{gaiadr3}, the Sloan Digital Sky Survey (SDSS; \citet{abdurro22}), Two Micron All Star (2MASS; \citet{2006AJ....131.1163S}), AKARI \citep{murakami2007}, and Wide-field Infrared Survey Explorer (WISE; \citet{wright2010}) passbands. We used BT-Settl (CIFIST) models \citep{baraffe15} to reproduce the spectral energy distribution (SED) of the star and to extrapolate to bluer and longer wavelengths. We obtained a luminosity of 0.748\,$\pm$\,0.010 L$_{\odot}$  for HD\,63433, which is very similar to previous results by \citet{mann20}. To estimate the radius and mass of the star, we followed the procedure presented in \citet{schweitzer19}. Based on the estimated effective temperature and luminosity and using the Stefan-Boltzman relation, we derived a radius of 0.934\,$\pm$\,0.029 R$_{\odot}$. Based on this radius and using the empirical mass-radius relations for solar type stars from \citet{eker18}, we determined a mass of 0.956\,$\pm$\,0.022 M$_{\odot}$, assuming that the star is on the main sequence, as expected for its age. A summary with the main stellar parameters of HD63433 can be found in Table\,\ref{tab:stellar_parameters}.

\begin{table}
\caption{Stellar parameters of HD\,63433\@.}
\label{tab:stellar_parameters}
\centering
\begin{tabular}{lcr}
\hline\hline
Parameter & Value & Reference\\
\hline
Name & HD\,63433 & HD\\
     & TIC 130181866  & TIC\\
     & TOI-1726  & --\\
\noalign{\smallskip}
$\alpha$ (J2000) & 07:49:55.1 & \textit{Gaia} DR3\\
$\delta$ (J2000) & 27:21:47.5 & \textit{Gaia} DR3\\
Sp.\ type & G5V & Gray03 \\ 
\noalign{\smallskip}
$\varpi$ [mas] & $44.685\pm0.023$ & \textit{Gaia} DR3\\
$d$ [pc] & $22.339\pm0.044$ & \textit{Gaia} DR3\\
RUWE & $0.991$ & \textit{Gaia} DR3\\
\noalign{\smallskip}
$T_{\text{eff}}$ [K] & $5553 \pm 56$ & This work\\
$\log{g}$ [cgs] & $4.56 \pm 0.08$ & This work\\
{[Fe/H]} [dex] & $-0.07 \pm 0.03$ & This work\\
\noalign{\smallskip}
$M_{\star}$ [M$_{\odot}$] & $0.956 \pm 0.022$ & This work\\
$R_{\star}$ [R$_{\odot}$] & $0.934 \pm 0.029$ & This work\\
$L_{\star}$ [L$_{\odot}$] & $0.748 \pm 0.010$ & This work\\ 
$v\sin i$ [km\,s$^{-1}$] & $7.26 \pm  0.15$ & This work\\
$v_{\mathrm{macro}}$ [km\,s$^{-1}$] & $2.73 \pm  0.15$ & This work\\
$P_{\mathrm{rot}}$ [d] & $6.4 \pm 0.4$ & This work\\
Ste.\ kinematic group & UMaG & Jon15 \\
Age [Myr] & $414 \pm 23$ & Jon15\\
$G$ [mag] & $6.737\pm 0.003$ & \textit{Gaia} DR3\\

\noalign{\smallskip}
\hline
\end{tabular}
\tablebib{
HD: \citet{cannon93};
TIC: \citet{Stassun2019};
\textit{Gaia} DR3: \cite{gaiadr3}; 
Gray03: \citet{gray03};
Jon15: \citet{jones15}
}
\end{table}

\subsection{Stellar rotation and activity}
\label{sec:prot}

\begin{figure}[ht!]
\includegraphics[width=1\linewidth]{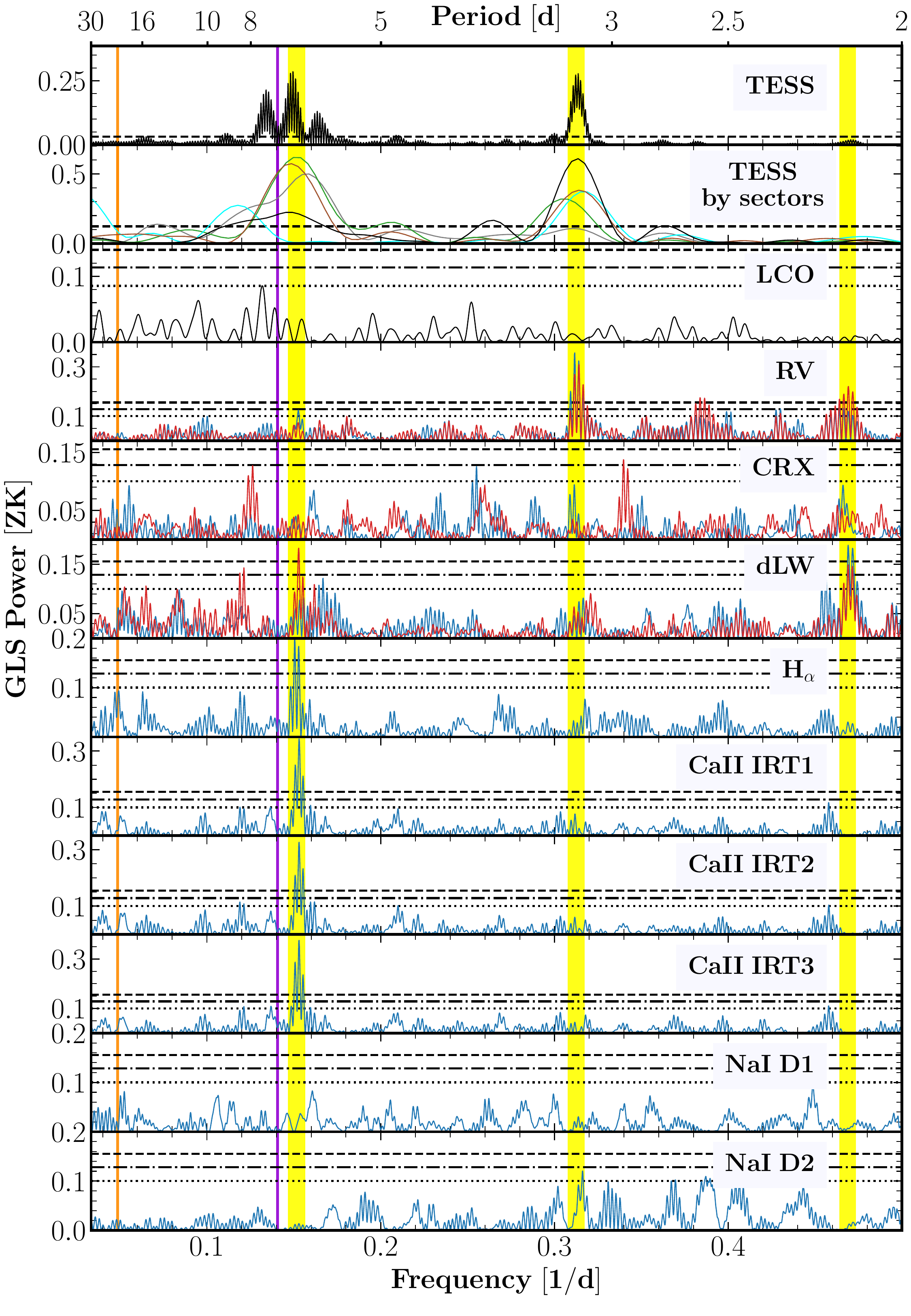}
\caption{GLS periodograms for the photometric light curves, RV, and spectral activity indicators from TESS, LCO, and CARMENES VIS and NIR data. The first panel shows the periodogram of the combination of all TESS sectors. A GLS periodogram for each sector (sectors 20, 44, 45, 46, and 47 depicted in grey, green, cyan, brown, and black, respectively) is shown in second panel. The third panel shows the periodogram of the LCO dataset as a solid black line. In the remaining panels, the periodograms calculated from CARMENES VIS data are displayed in blue, and those calculated for the NIR arm are plotted in red. In all panels, the solid purple and vertical orange lines indicate the orbital periods of planets b and c, respectively. The shaded yellow bands indicate the rotation period (6.4d), half (3.2d), and one-third of the rotation period (2.1d). The dashed horizontal black lines correspond to FAP levels of 10\%, 1\%, and 0.1\% (from bottom to top). The first two panels only include the 0.1\% FAP multiplied by a factor 100 for clarity.
\label{fig:act_ind}}
\end{figure}

As a consequence of the youth and fast rotation of HD\,63433, the star shows a remarkable stellar activity that is clearly visible in photometry as well as in RV due to the rotation of the star. HD\,63433 has a stellar rotation period of 6.45\,$\pm$\,0.05 days \citep{mann20}, as estimated from the combination of the Lomb-Scargle periodogram and the autocorrelation function of the TESS light curve in sector 20\@. We searched for periodic signals between 2 and 30 days using the larger temporal coverage of sectors 44--47 (more than 100 days of quasi-continuous observations), our dedicated photometric monitoring with LCO, and spectroscopic campaigns with the CARMENES instrument, using the RV and the stellar activity indicators that are generated by \texttt{serval} for the CARMENES VIS and NIR datasets. GLS periodograms \citep{GLS} were computed for these data and are shown in Fig.\,\ref{fig:act_ind}. The top two panels show the GLS periodograms of TESS for all the sectors and of each one of them, respectively. The third panel shows the LCO periodogram, and the following panels present the RV periodograms and different activity indicators for both the VIS and NIR. 
The periodograms of the TESS, dLW (NIR), H$\alpha$, and \ion{Ca}{II} indexes show a period between 6.4--6.8 days as the most significant signal, in agreement with the published rotation period \citep{mann20}. On the other hand, the RV periodogram shows that the strongest signal is at $\sim$3.2 days, that is, at\ half the rotation period. This signal is also very intense and comparable with the 6.4- to 6.8-day signal in the TESS periodogram. It is easily recognizable in Fig.\,\ref{fig:LC_TESS}. The signal of one-third of the rotation period (centred on 2.1 days) is also detected in the dLW (VIS) panel, in the RV, and the dLW (NIR) periodogram, but with a lower significance. Finally, the LCO, CRX, and \ion{Na}{I} indicators do not show any significant signal. The photometric precision obtained with the LCO data in the $V$ band ($\sim$17-46 mmag) was not sufficient to detect the photometric variations of the star, which show a peak to peak of $\sim$20 mmag in TESS data, partially because there were no stars of similar brightness in the field of view and because the chromaticity of these activity signals is low. For this reason, and because no significant signals are actually observed in the periodogram, we decided not to use these data in the subsequent analysis. We conclude that the signal of the stellar rotation period is present (in its fundamental frequency and/or its harmonics) in most of our data. However, it is remarkable that the most significant peaks in the TESS periodograms (probably the best indicator, considering the cadence and baseline of the data) have a similar intensity and a FWHM of $\sim$0.4 days. The second panel shows that the peak values change slightly from one sector to the next and between the fundamental signal and its first harmonic. This may be an indication of variations in the rotation period determination that may be caused by differential rotation or by variations in the activity regions on timescales of a few rotation periods. Lastly, to study the correlation between the CARMENES data products, we computed the Pearson $r$ coefficient. We found correlations of $p$\,$>$\,0.8 between RV VIS and RV NIR on the one hand and between the \ion{Ca}{II} indices on the other hand. We found no evidence of a correlation for the other products.

In Sects.\,\ref{sec:tr}, \ref{sec:rv}, and \ref{sec:joint} we use Gaussian process regression (GP; \citealp{Rasmussen2006}), which is implemented in the \texttt{celerite} \citep{celerite} and \texttt{George} \citep{ambikasaran15} packages, to derive the hyperparameters of stellar activity, including the parameter associated with the period. Using the data from TESS, CARMENES VIS and NIR, we measured a $P_{\mathrm{rot}}$\,=\,6.40\,$\pm$\,0.02\,d (see Sect.\,\ref{sec:joint} and Table \ref{tab:joint-fit}), consistent with the values analysed in the periodograms and with previous results. For all the considerations given above, we adopted a final rotation period of 6.4\,$\pm$\,0.4\,d, where we used the FWHM of the main peak of the TESS periodogram as the error (Fig.\,\ref{fig:act_ind}, top panel).

\section{Transit analysis}
\label{sec:tr}

\cite{mann20} identified four transits of HD\,63433\,b and two transits of HD\,63433\,c in TESS sector 20\@. We employed the new TESS sectors (44, 45, 46, and 47) to update the planetary transit parameters and to search for additional new planets, taking advantage of more than 100 days of continuous observations. This allowed us to identify planets with longer orbital periods. However, as shown in Figs.\ \ref{fig:LC_TESS} and \ref{fig:RV_CARM}, young stars are dominated by intense periodic stellar activity with timescales of several days. This represents the main difficulties in finding young planets. Therefore, it is necessary to find the appropriate model to capture this type of activity. We chose GPs as an effective model that is flexible enough to model the variations of the amplitudes and the quasi-periodic behaviour that this stellar activity shows. The activity may show changes between two periods and even within the same cycle \citep{angus18}. In particular, to model the photometric stellar activity, following a similar kernel as in \cite{masca21}, we used the double simple harmonic oscillator (dSHO), built as the sum of two SHO kernels, defined as 

\begin{equation}
\begin{aligned}
k_{\mathrm{dSHO}}(\tau) = & \  k_{\mathrm{SHO}}(\tau; \eta_{\sigma_{1}}, \eta_{L_{1}}, \eta_{P}) + k_{\mathrm{SHO}}(\tau; \eta_{\sigma_{2}}, \eta_{L_{2}}, \eta_{P}/2) \\
               = & \ \eta_{\sigma_{1}}^{2}  e^{-\frac{\tau}{\eta_{L_{1}}}}  \left[ \cos \left(\eta_{1} \frac{2\pi\tau}{\eta_{P}} \right) + \eta_{1} \frac{\eta_{P}}{2\pi \eta_{L_{1}}} \sin \left(\eta_{1} \frac{2\pi\tau}{\eta_{P}} \right) \right] \\
             & + \eta_{\sigma_{2}}^{2}  e^{-\frac{\tau}{\eta_{L_{2}}}}  \left[ \cos \left(\eta_{2} \frac{4\pi\tau}{\eta_{P}} \right) + \eta_{2} \frac{\eta_{P}}{4\pi \eta_{L_{2}}} \sin \left(\eta_{2} \frac{4\pi\tau}{\eta_{P}} \right) \right],
\end{aligned}
\label{eq:dsho}
\end{equation}

 where $\tau$\,$\equiv$\,$|t_i - t_j|$ is the time difference between two data points, $\eta$\,$\equiv$\,$|1 - (2\pi \eta_{L}/\eta_{P})^{-2} |)^{1/2}$ and $\eta_{\sigma_{i}}$, $\eta_{L_{i}}$, and $\eta_{P}$ are the hyperparameters that represent the amplitude of the covariance, the decay timescale, and the period of the fundamental signal, respectively. The induced stellar activity can act on different timescales, from hours or days (oscillations and granulation), and from days to months (rotation) or years (magnetic cycles). The hyperparameters of the GPs allowed us to adapt the model to the scales that we wished to fit. In our case, these were the variations due to stellar rotation. Therefore, we interpret the period of the process physically as the rotation period of the star, the amplitude of the covariance as the amplitude variations produced by the rotation of a particular configuration of the active regions, and the decay timescale as the evolution of these active regions of the star. The definition of this kernel is valid as long as $\eta_{P}$\,$<$\,$2\pi \eta_{L}$ , which it is a reasonable assumption in young stars because it is observed that a clearly periodic behaviour dominates in the activity. This kernel has been widely used in the literature \citep{david18a, mann20, newton21, toff21} to model the light curves of young stars, for which the rotation period and its first harmonic is easily identifiable. This is consistent with our periodogram (TESS panel in Fig.\,\ref{fig:act_ind}), and when there is a sufficient cadence of data because it allows great flexibility. An instrumental offset ($\gamma_{\mathrm{TESS}}$) was also considered for the TESS dataset, and a jitter ($\sigma_{\mathrm{jit,TESS}}$) term was added in quadrature to the error bars. To explore the parameter space in our analysis, we employed an affine-invariant ensemble sampler \citep{goodman10} for the Markov chain Monte Carlo (MCMC) implemented in the \texttt{emcee} code. The parameter space here was also explored with another sampler algorithm, \texttt{dynesty}\footnote{\url{https://github.com/joshspeagle/dynesty}} \citep{dynesty}, which is based on nested sampling \citep{skilling04}. The next sections show that the results are consistent.

\subsection{Transit search}
\label{sec:tr_search}

We searched for transits in the TESS light curves using the box least-squares periodogram  \citep[BLS;][]{kovacs2002,hartman2016} in each individual sector, but also in the combination of all of them. We modelled the stellar activity and subtracted it as mentioned above, and then applied the BLS to search for transits. To do this, we forced (setting normal priors) the amplitude covariance hyperparameters ($\eta_{\sigma_{1}}$, $\eta_{\sigma_{2}}$) to values close to the standard deviation of the light curve ($\sim$4.3 ppt) and fixed the $\sigma_{\mathrm{jit,TESS}}$ parameter to $\sim$0.8 ppt, which corresponds to the depth of the largest transit of the two known planets (HD\,63433\,c). In this way, we modelled the stellar activity smoothly and prevented the GPs from removing or modelling possible transits.

After the stellar activity was subtracted, BLS found HD\,63443\,b as the most significant signal with 18 transits (vertical orange bands in Fig.\,\ref{fig:LC_TESS}). Then, we masked out this signal and iteratively applied  the BLS algorithm again to search for additional features. It found 8 transits of HD\,63433\,c as the second strongest signal (vertical purple bands in Fig.\,\ref{fig:LC_TESS}). After this, we found no additional transit. Furthermore, we inspected the light curves by eye for isolated transits and found no variations compatible with transits.

\subsection{Transit parameters}
\label{sec:trfit}

To improve the ephemerides of the two known planets through a larger baseline, we proceeded to simultaneously fit our photometric model as a combination of stellar activity and two planetary transit signals. To model the planetary transits, we used \texttt{PyTransit}\footnote{\url{https://github.com/hpparvi/PyTransit}} \citep{Parviainen2015}, and the stellar activity was modelled in the same way as in Sect.\ \ref{sec:tr_search}, now leaving $\sigma_{\mathrm{jit,TESS}}$ as a free parameter. Our transit-only fit contains the following planetary parameters: the time-of-transit centre ($T_c$), the planetary radius ($R_p$), the orbital period of the planet ($P$), and the impact parameter ($b$). Models with non-circular orbits were also included. The eccentricity ($e$) and argument of periastron ($\omega$) in them were parametrized as in \cite{anderson2011}($\sqrt{e}\sin{\omega}$, $\sqrt{e}\cos{\omega}$). The input for stellar parameters was the stellar mass ($M_\star$), the stellar radius ($R_\star$), and the linear and quadratic limb-darkening coefficients ($q_1$, $q_2$). In the last case, we used the parametrization of \cite{kipping02013}, where the initial values were previously calculated with the \texttt{Python Limb Darkening Toolkit}\footnote{\url{https://github.com/hpparvi/ldtk}} (PyLDTk; \citealp{Parviainen2015b}). We chose to sample the hyperparameters on a natural logarithmic scale for faster convergence. In addition, we partially constrained some of them assuming the following hypotheses: As mentioned before, the standard deviations of the process were favoured by setting normal priors to the scales of the standard deviation of the activity, the decay timescale was limited to values clearly greater than the rotation period and lower than the baseline of the data sample, and finally, the period of the process was allowed to sample between the main signals of activity (6.4d, 3.2d, and 2.1d). The priors and posterior results are presented in Table \ref{tab:tronly-fit}, assuming circular and eccentric orbits for the planets. Our posterior parameters are consistent with those obtained by \cite{mann20}.

\subsection{Transit-timing variations}

\begin{figure}[ht!]
\includegraphics[width=1\linewidth]{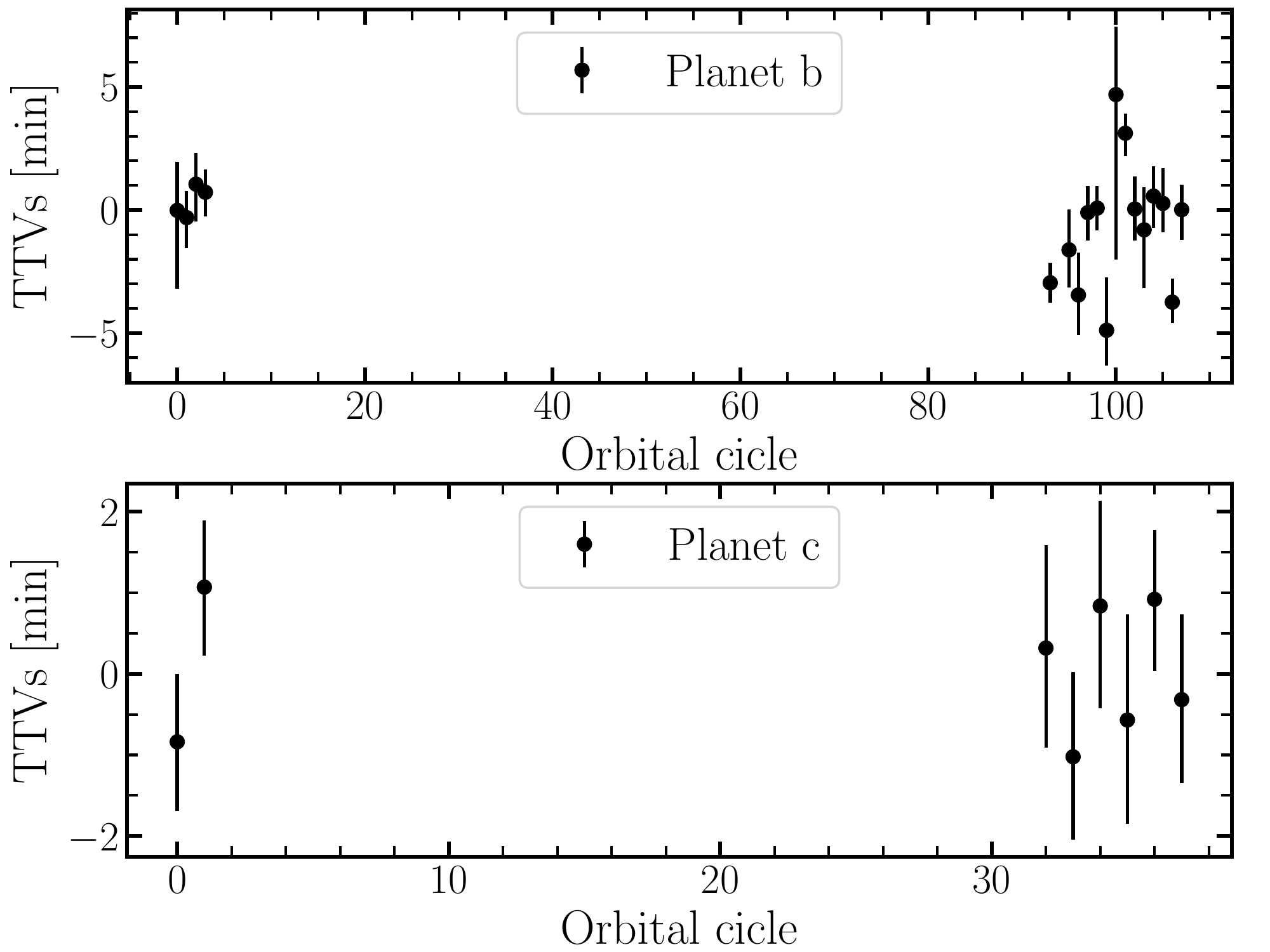}
\caption{TTVs for planet b (top panel) and planet c (bottom panel) represented as black dots with their 1$\sigma$ uncertainty.
\label{fig:TTV}}
\end{figure}

Because we have a long temporal coverage ($\sim$750\,d) and 18 and 8 transits for planet b and c, respectively, the transit-timing variations (TTVs) can be a tool for measuring the planetary masses. To do this, we proceeded in the same way as in the previous section, but used  $T_{c}$ for each individual transit. Fig.\,\ref{fig:TTV} shows that the TTVs of planet b and c present variations consistent with zero within the error bars. Therefore, we conclude that no TTVs are detected in the system to measure planetary masses.

\section{RV analysis}
\label{sec:rv}

With the new ephemeris of the planets derived in the previous section, we checked whether some RV epochs fall during the transits because these RVs may be altered due to the RM effect. Three RV measurements fall during the transit of HD\,63433\,b (BJD\,=\,2459513.5797, 2459634.4229) and HD\,63433\,c (BJD\,=\,2459131.6311). Because the expected variation in amplitude due to RM effect of HD\,63433\,b and c can be a few m\,s$^{-1}$ \citep{fei20, zhang22}, we decided to remove these RVs from our data. Overall, our final dataset contains 147 RV points of CARMENES VIS and 143 of CARMENES NIR\@.

\subsection{Periodogram analysis}
\label{sec:rv-gls}

\begin{figure}[ht!]
\includegraphics[width=1\linewidth]{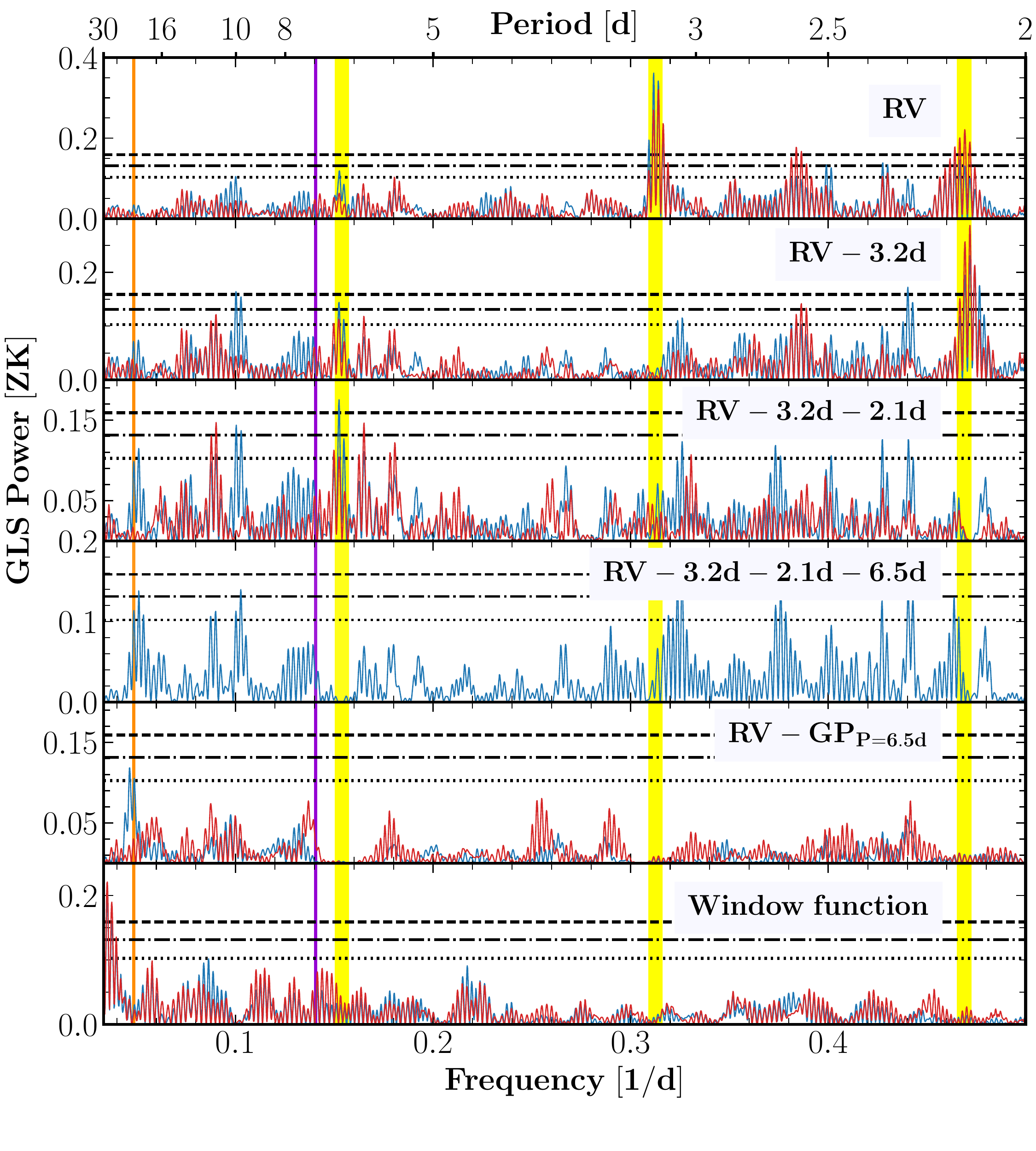}
\caption{GLS periodograms for the CARMENES VIS (blue) and NIR (red) RV datasets. In all panels, the solid vertical purple and orange lines indicate the orbital periods of planets b and c, respectively. The shaded vertical yellow bands indicate the rotation period (6.4d) and the half (3.2d) and one-third (2.1d) harmonics. The dashed, dotted-dashed, and dotted horizontal black lines correspond to FAP levels of 10\%, 1\%, and 0.1\% from bottom to top, respectively.
\label{fig:rv_prewhite}}
\end{figure}

In Fig.\,\ref{fig:rv_prewhite} we show a more detailed study of the RV periodograms calculated in Sect.\,\ref{sec:prot}. The CARMENES VIS and NIR datasets are highlighted in blue and red, respectively. The orbital periods of the transiting planets b and c are marked with vertical purple and orange lines, respectively. In the top panel in Fig.\,\ref{fig:rv_prewhite}, we plot the RV data, where two peaks with a high significance (FAP<0.1\%) are seen at $\sim$3.2 and $\sim$2.1 days. They correspond to the harmonics of the stellar rotation period (vertical yellow band) estimated in Sect.\,\ref{sec:prot}. As previously commented, the signal of the planets that orbit or transit young stars is usually hidden by the strong stellar activity, which is the dominating signal in periodograms. As a first approximation, we modelled the most significant signals with a sinusoidal function before subtraction. The second panel shows the residuals after the signal related to half of the stellar rotation period was subtracted. This is the most significant signal in both periodograms. After removal of this activity signal, the 2.1d signal becomes the most significant in both datasets, while the signals of the planets remain hidden. The data from both arms seem to be dominated by stellar activity so far, with a strong correlation between the two datasets (Sect.\,\ref{sec:prot}). In the third panel, we repeat the process and subtract the 2.1d signal. In the periodogram of the CARMENES VIS residuals, only one signal remains with an FAP<0.1\%: the signal at 6.4d. This signal is\ at the rotation period of the star. However, in the CARMENES NIR dataset, all the peaks have a significance lower than 0.1\% FAP. In the fourth panel, we subtracted the three most significant signals associated with stellar activity only in the CARMENES VIS data. The signal with an FAP$\sim$1\% close to the orbital period of planet c does not appear in the periodogram of CARMENES NIR because the larger uncertainties in the NIR measurements make it more difficult to detect signals with smaller amplitudes. In the fifth panel, we subtracted the stellar activity for each dataset by modelling in another way, using GP (see Sect.\,\ref{sec:rvfit}). As with the previous method,  the signal of the outer planet (FAP$\sim$10--1\%) is visible, but not that of the inner one. The orbital period of HD\,63433\,b is $\sim$7.1d and the rotational period is $\sim$6.4d, that is, within a 10\% margin. This quasi-coupling of the signals, together with the great difference in amplitude between them, makes it especially difficult to detect the planetary signal for the inner planet. Finally, in the sixth panel, we show the window functions of the VIS and NIR datasets. In summary, from the analysis of the periodograms of the VIS and NIR RV, we conclude that only the activity signal in both datasets and only the outer planet signal in CARMENES VIS is detected with sufficient significance.

\subsection{RV modeling}
\label{sec:rvfit}

After studying the signals present in the periodograms, we tested a battery of RV-only fits to determine the best activity model and the best Keplerian model of the planets. We considered two possible datasets: only the VIS dataset, and the combination of VIS and NIR measurements. To create a model of stellar activity, we used GP as in Sect.\ \ref{sec:trfit} and the quasi-periodic (QP) kernel defined in \cite{aigrain12} as

\begin{equation}
k_{\mathrm{QP}}(\tau)  =  \eta_{\sigma}^2 \exp \left[ -\frac{\tau^2}{2\eta_{L}^2} -\frac{\sin^2{ \left( \frac{\pi \tau}{\eta_{P}} \right)}}{2\eta_{\omega}^2} \right], \\
\label{eq:qp}
\end{equation}

where $\tau$, $\eta_{\sigma}$, $\eta_{L}$, and $\eta_{P}$ are the same as in Eq.\,\ref{eq:dsho}, and $\eta_{\omega}$ controls the weight given to the periodic part of the kernel. The QP kernel has one free parameter less than the dSHO, and there is no coupling between the periodic part and non-periodic part, as in the case of dSHO, which gives it less flexibility when the data sampling is smaller, as is the case with RV data. Once again, we imposed normal priors on the covariance amplitudes associated with the standard deviation of the data because the goal was to model stellar activity with a smooth function on period scales. As the rotation period signal and its harmonics are present in RV data, we also tested a combination of sinusoidal functions to fit the stellar activity. To do this, we considered several cases. The first case was an activity-only model, for which we assumed that planetary signals are not detected in the CARMENES RV data. We modelled the stellar activity with GPs or sinusoidal functions. The second case considered the existence of one transiting planetary signal (the inner or the outer planet) modelled as a Keplerian circular orbit with the initial planetary parameters obtained from a transit-only fit (Sect.\,\ref{sec:trfit}), including the stellar activity as in the first case. The last case included a Keplerian model of both planets and the different models for stellar activity. The Keplerian model parameters were the $T_c$, $P$, and the RV amplitude of the planet ($K$) and were modelled with the \texttt{RadVel}\footnote{\url{https://github.com/California-Planet-Search/radvel}} \citep{fultonradvel} package. In the photometric analysis, all models of the RV fit included an instrumental offset ($\gamma_{\mathrm{CARMENES}}$) and a jitter term added in quadrature to the error-bar measurements ($\sigma_{\mathrm{jit,CARMENES}}$).

To evaluate which model was better for each dataset, we considered the Bayesian log-evidence ($\ln \mathcal{Z}$) criterion defined in \cite{trotta2008} and calculated from \cite{diaz16}, where the model with a higher log-evidence is favoured and was used as reference ($\Delta \ln \mathcal{Z}$\,$\equiv$\,$\ln \mathcal{Z}_{model} - \ln \mathcal{Z}_{model\ ref}$). When the difference between two models is $|\Delta \ln \mathcal{Z}|$\,$<$\,1, the two models are considered indistinguishable. When 1 $<$\,$|\Delta \ln \mathcal{Z}|$\,$<$\,2.5, the evidence in favour of one of the model is weak, and when the difference is 2.5 $<$\,$|\Delta \ln \mathcal{Z}|$\,$<$\,5, the evidence is moderate. This criterion allows us to compare models with different numbers of parameters as long as the input observational dataset is the same. The results of the different models considered in this work are compiled in Table \ref{tab:logZ}. For the CARMENES VIS dataset, the model with the best log-evidence is always the model that includes the a single planetary model, planet c (marked in bold in the table). This yields weak evidence compared to other planetary models. In contrast, including a model with planet b alone does not give a significant improvement. Moreover, better $\ln$\,$\mathcal{Z}$ are clearly obtained when the stellar activity is modelled by GPs instead of sinusoidals. The amplitudes obtained for the planets using GPs or sinusoidal functions are comparable within 1$\sigma$, although those obtained for planet c using GPs have a slightly lower value.

\begin{table*}[htbp]
\begin{center}
\begin{tabular}{cc|ccc|ccc}
\hhline{~~======}
 & & \multicolumn{6}{c}{Activity model}\\
\cline{3-8}
 & & \multicolumn{3}{c}{GP (QP Kernel)} & \multicolumn{3}{|c}{3 Sin (P$_{1}\sim$6.4d, P$_{2}\sim$3.2d, P$_{3}\sim$2.1d)}\\
\hline
Dataset & Planets & K$^b$[m\,s$^{-1}$] & K$^c$[m\,s$^{-1}$] & $\Delta \ln \mathcal{Z} $ & K$^b$[m\,s$^{-1}$] & K$^c$[m\,s$^{-1}$] & $\Delta \ln \mathcal{Z}$\\
\hline
VIS & 0 & -- & -- & $-$2.2 & -- & -- & $-$41.2\\
VIS & 1 & 3.42\,$\pm$\,2.95 & -- & $-$4.2 & 1.89\,$\pm$\,1.76 & -- & $-$41.5\\
VIS & 1 & -- & \textbf{3.72\,$\pm$\,1.12} & \textbf{0} & -- & 6.13\,$\pm$\,1.69 & $-$37.8\\
VIS & 2 & 2.68\,$\pm$\,2.51 & 3.62\,$\pm$\,1.11 & $-$1.6 & 1.86\,$\pm$\,1.55 & 6.30\,$\pm$\,1.70 & $-$42.0\\
\hline
VIS+NIR & 0 & -- & -- & $-$3.8 & -- & -- & --\\
VIS+NIR & 1 & 2.20\,$\pm$\,2.05 & -- & $-$6.8 & -- & -- & --\\
VIS+NIR & 1 & -- & 3.77\,$\pm$\,0.97 & 0 & -- & -- & --\\
VIS+NIR & 2 & \textbf{1.77 \,$\pm$\,1.76} & \textbf{3.69\,$\pm$\,0.98} & \textbf{$-$2.1} & -- & -- & --\\
\hhline{========}
\end{tabular}
\caption{Model comparison for the RV-only analysis of HD\,63433 using the difference between log-evidence ($\Delta \ln \mathcal{Z}$). In the model name, "3 Sin" refers to three sinusoidal functions and its periods. All models assume circular orbits, and the amplitudes are given with their 1$\sigma$ uncertainty.}
\label{tab:logZ}
\end{center}
\end{table*}

In the second configuration, using both CARMENES VIS and NIR data, we only ran the activity model with GP, which has the best $\ln$\,$\mathcal{Z}$ by far. Although the error bars in CARMENES NIR data are significantly higher, the stellar activity appears to be well detected in the periodograms. Therefore, a combined model of both datasets may help to better constrain stellar activity and, consequently, to better define planetary signals. Since there is a high correlation between VIS and NIR datasets, we have shared the hyper-parameters of $\eta_{L}$, $\eta_{\omega}$, and $\eta_{P}$ between both datasets in our model. At the bottom of Table \ref{tab:logZ}, we see that the best model includes planet c, together with the activity, in the same way as with CARMENES VIS data alone. In this case, the amplitudes are similar  to the previous case, but the errors are slightly lower. After this analysis, we decided to use the results that include CARMENES VIS and NIR data with two planetary models for two reasons. Although we do not see any signal that we can associate with the planets in the periodogram of CARMENES NIR, the stellar activity seems to be well detected. A better characterization of stellar activity (the hyperparameters of the model that combines the two datasets are slightly better constrained than when CARMENES VIS is used alone) appear to constrain the planetary parameters better. Finally, we also decided to adopt the two-planet models as the final model because we have a priori information of the existence of the two transiting planets, and with this model, we can place a constraint on the mass of planet b. The planetary parameters for this best RV-only fit are catalogued in Table \ref{tab:rvonly-fit} for circular and eccentric orbits.

\subsection{Injection-recovery tests of planet b}
\label{sec:rv-pl_b}

\begin{figure}[ht!]
\includegraphics[width=1\linewidth]{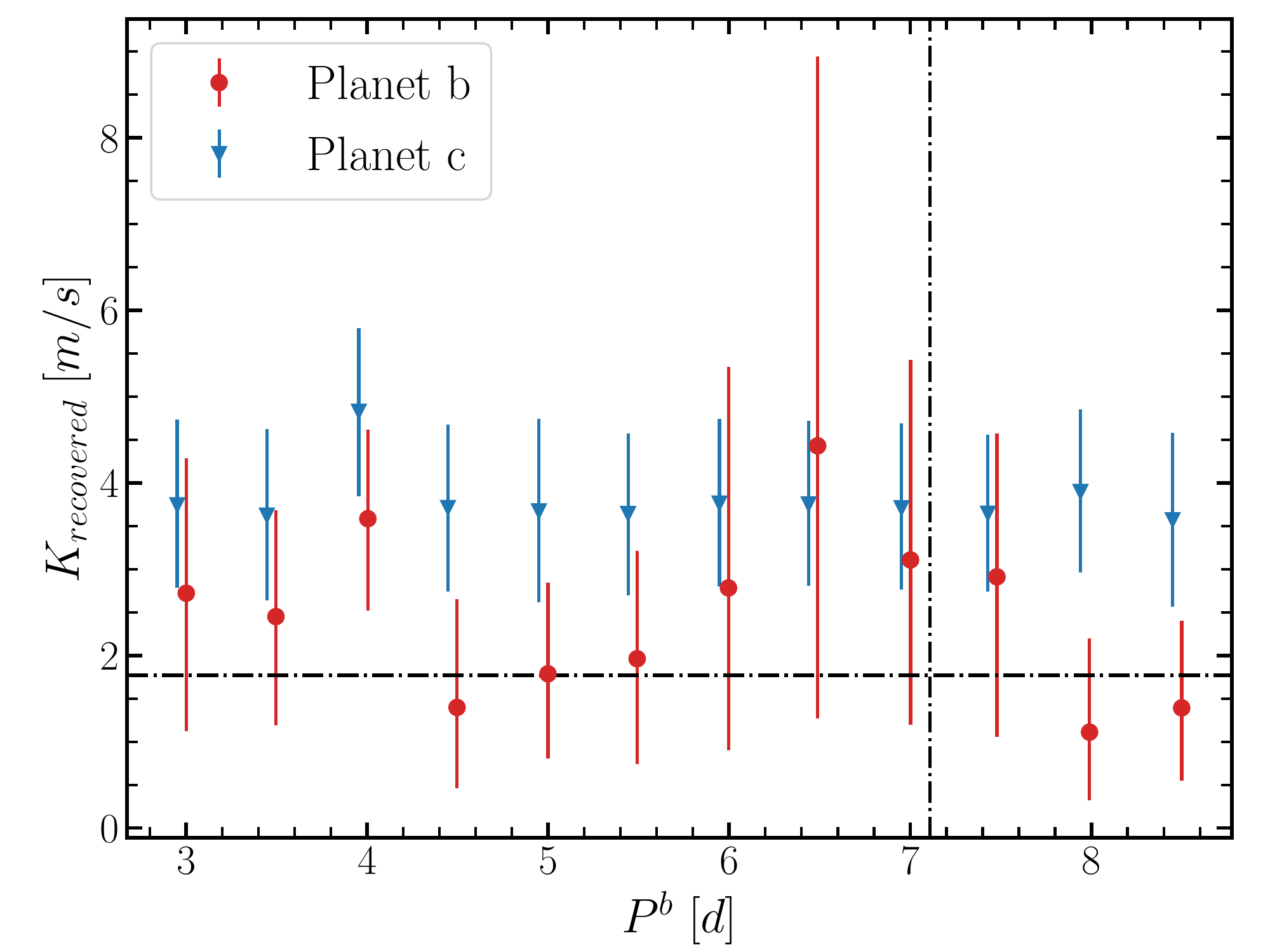}
\caption{Recovered RV amplitudes vs. orbital periods for planet b. The red dots and blue triangles show the amplitude we recovered for planets b and c, respectively. The dotted horizontal and vertical black lines indicate the injected amplitude and the orbital period for planet b, respectively..
\label{fig:rv-pl_b}}
\end{figure}

As mentioned in Sect. \ref{sec:rv-gls}, the proximity of the signals of the stellar rotation and the orbital period of the inner planet hamper a correct measurement of the RV amplitude of planet b. In order to test the dependence of the amplitude of planet b on the separation between the orbital and rotation period, we decided to carry out a injection-recovery test. First, we subtracted the signal of the inner planet obtained from the best RV model that combines GP and two Keplerians ($K^b$\,$\sim$\,1.77 m\,s$^{-1}$). Then, we injected the signal of planet b with the same amplitude but slightly different orbital periods, and finally, we determined the amplitude of the inner planet using the same model as in Sect. \ref{sec:rvfit}. In Fig. \ref{fig:rv-pl_b} we represent the recovered amplitude for different orbital periods. For values close to the rotation period and its first harmonic, the recovered amplitude and the error bars can be twice or three times as high as the value of the injected amplitude. This confirms the dependence of the orbital and rotation periods. In addition, it is noteworthy that even for the orbital periods farthest from the rotation period, the uncertainty of the amplitude is never lower than $\sim$0.8 m\,s$^{-1}$, indicating that with this dataset and model, amplitudes smaller than 2.4 m\,s$^{-1}$ could not be measured with a 3$\sigma$ significance. The figure also shows that the amplitude measure of the outer planet depends on the orbital period of the inner planet. The amplitude obtained for planet c is displayed as blue triangles (and intentionally shifted to the left of each measure of planet b) to show that it is stable and independent of the measure of planet b.

\section{Joint-fit analysis}
\label{sec:joint}

\begin{figure*}[ht!]
\includegraphics[width=0.49\linewidth]{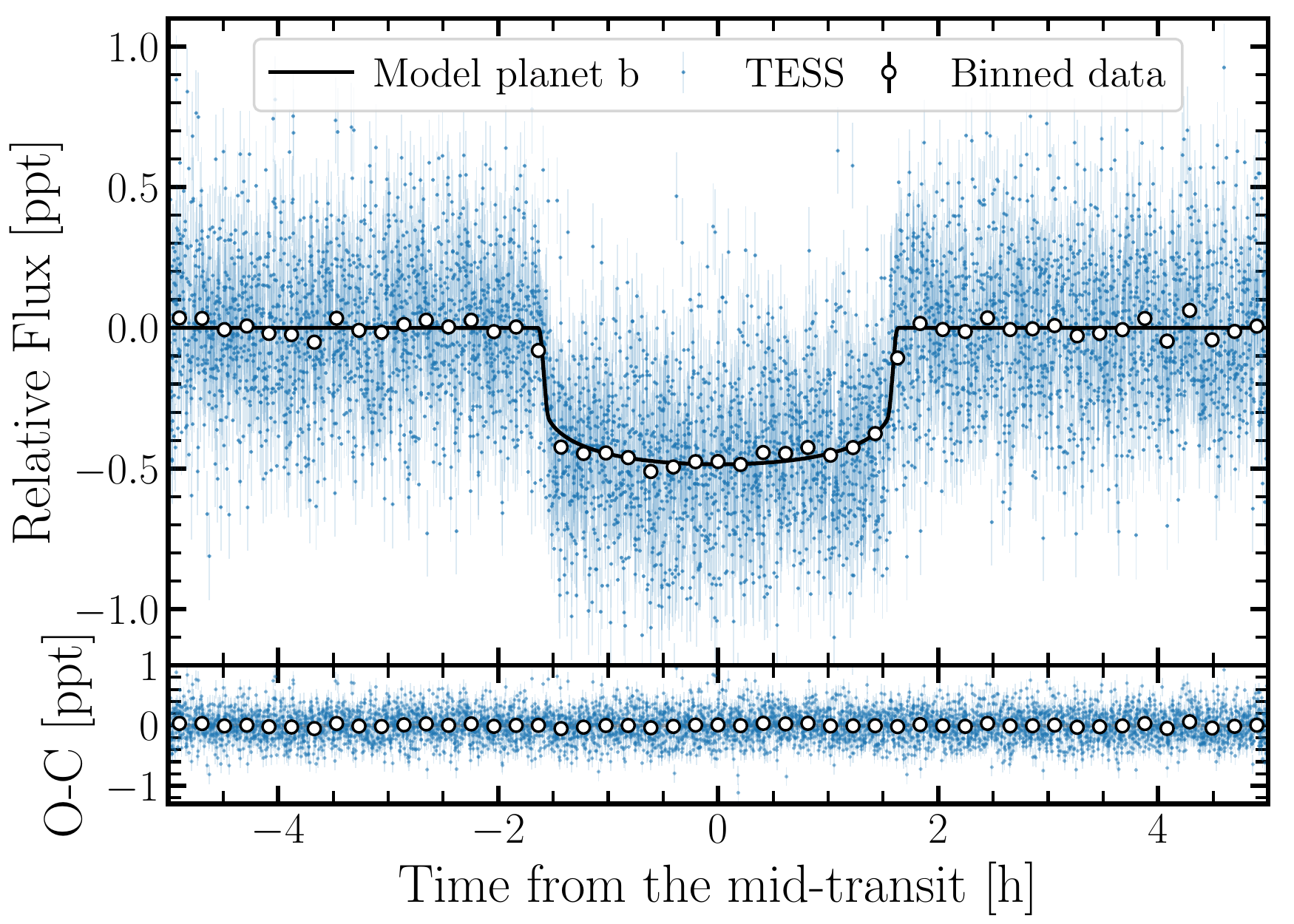}
\includegraphics[width=0.49\linewidth]{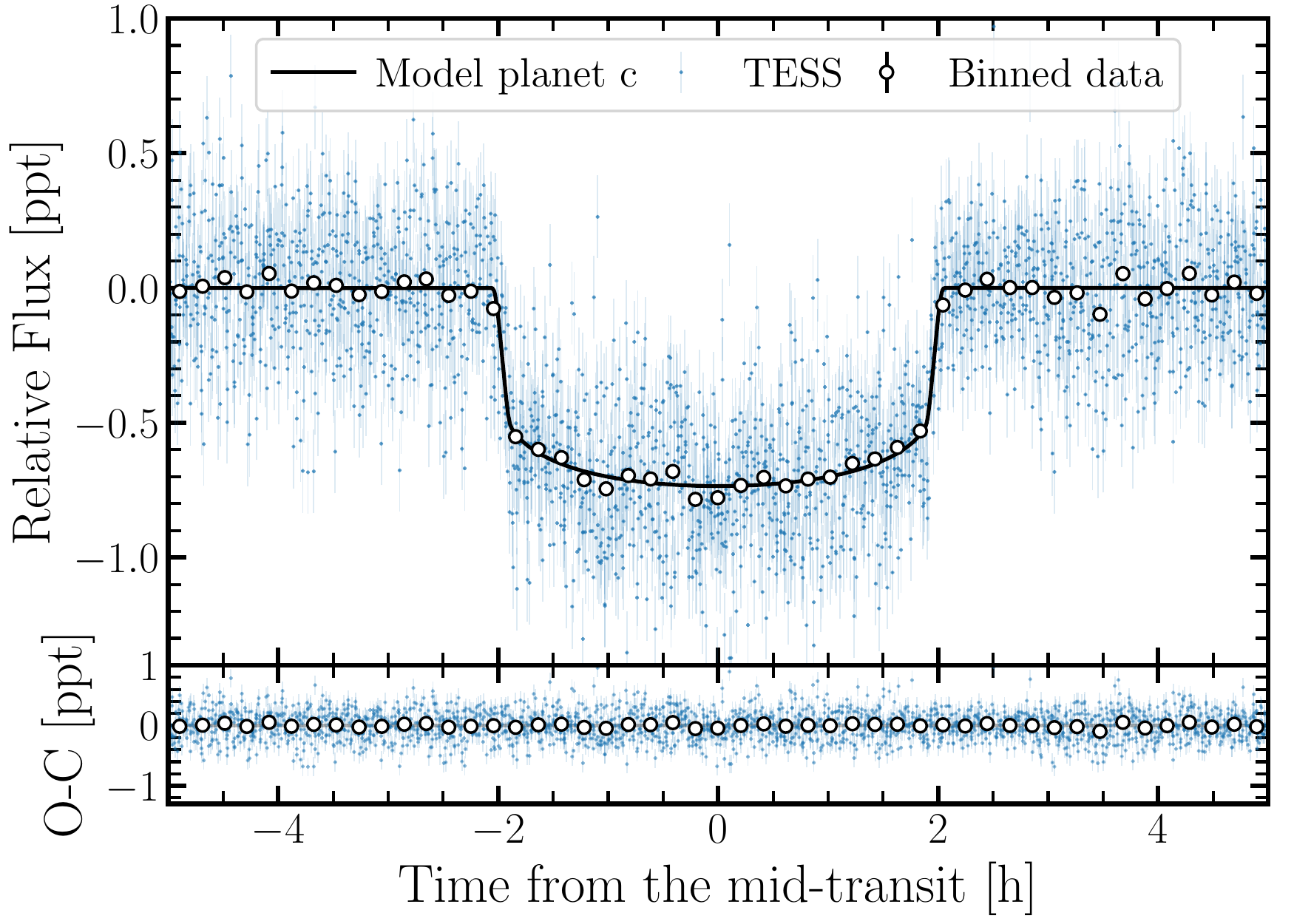}
\caption{The top panels show the phase-folded TESS light curves of the transits of planets b and c (blue points), binned data (white dots), and our best transit-fit model (black line). The bottom panels show the residuals for the best fit.  
\label{fig:tr_folded}}
\end{figure*}

\begin{figure*}[ht!]
\includegraphics[width=0.49\linewidth]{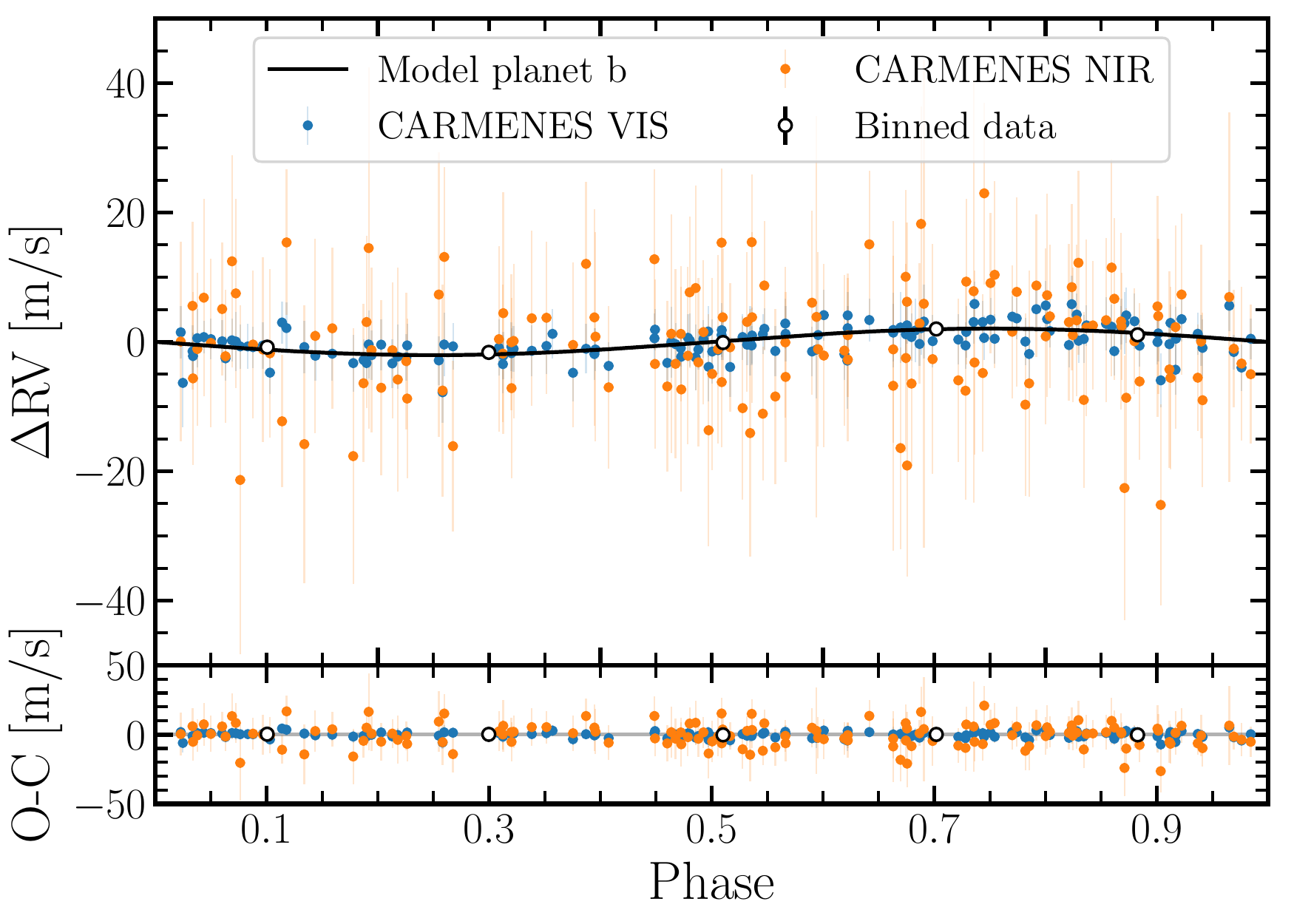}
\includegraphics[width=0.49\linewidth]{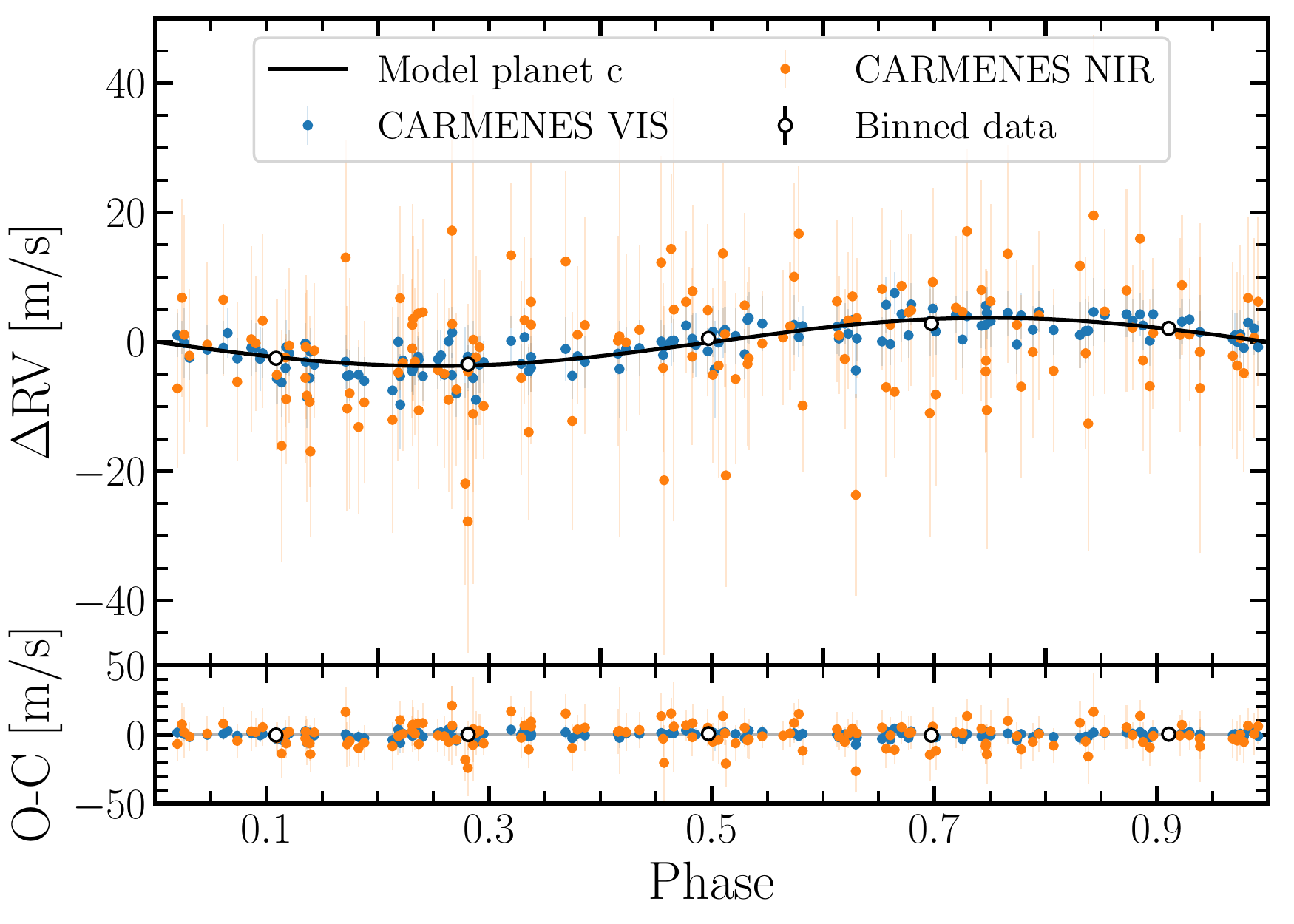}
\includegraphics[width=0.49\linewidth]{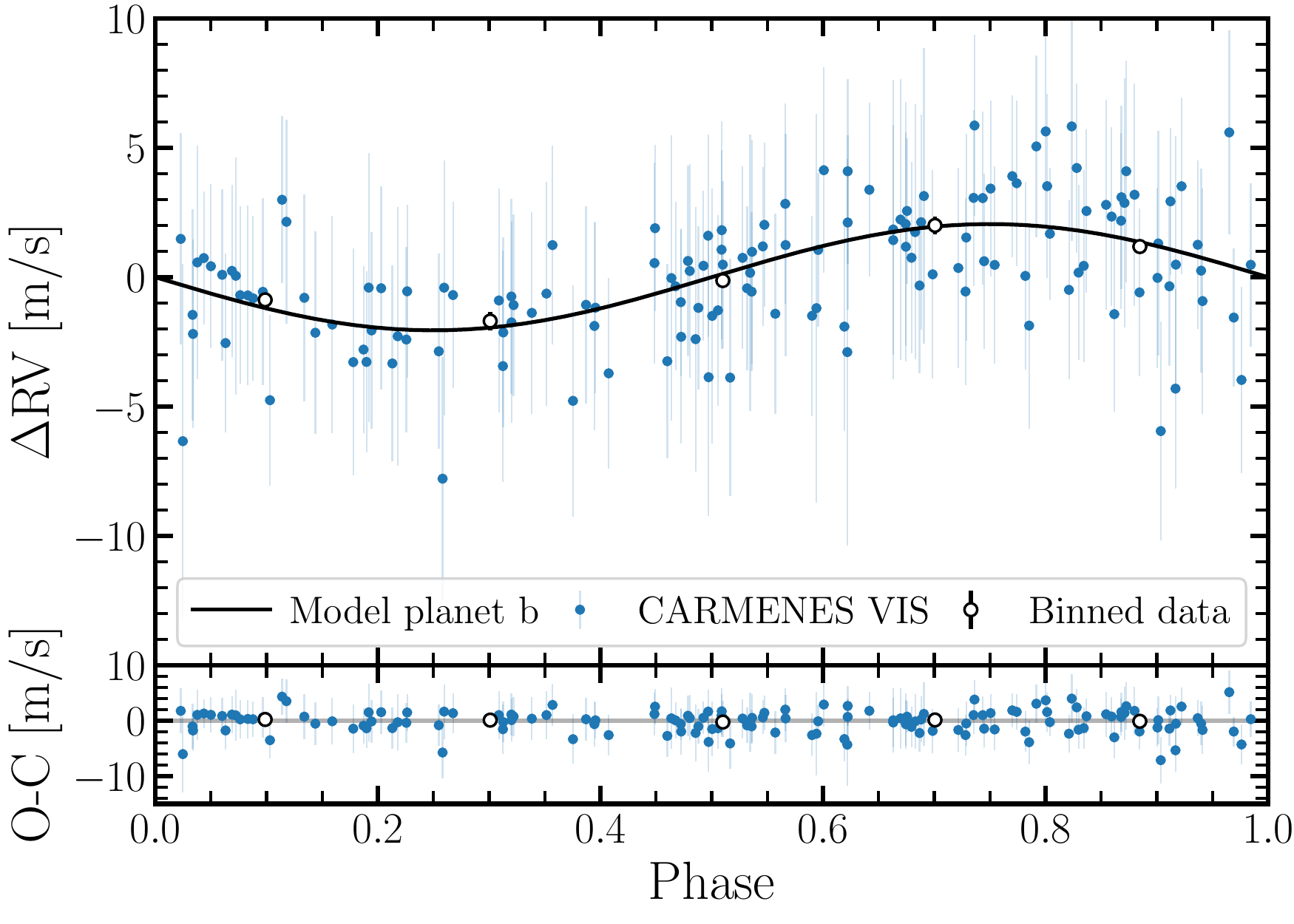}
\includegraphics[width=0.49\linewidth]{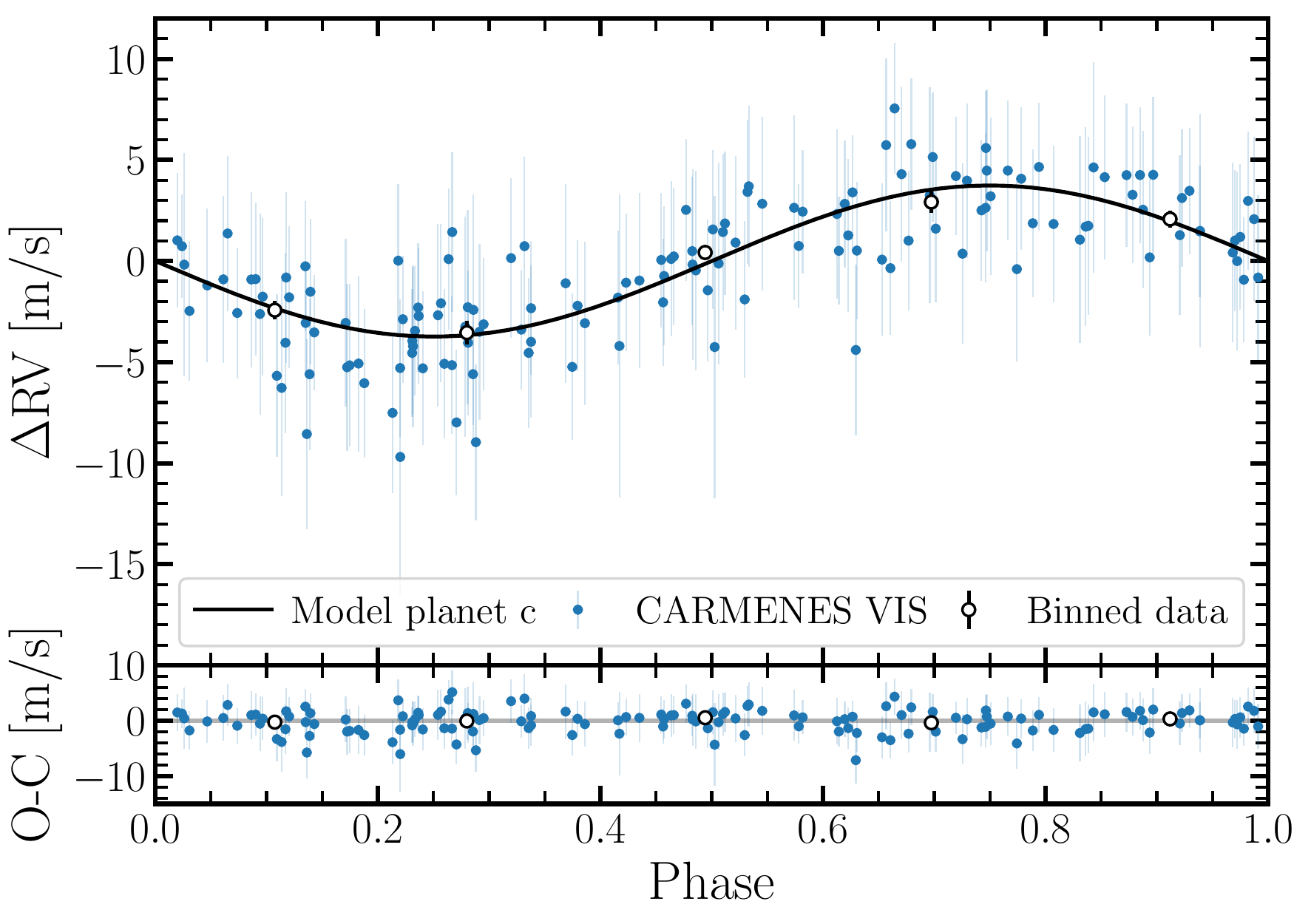}
\caption{The top panels show RV-folded CARMENES data for planets b and c (blue and orange dots for VIS and NIR, respectively), binned data (white dots), and our best Keplerian-fit model (black line). The bottom panels show the residuals for the best fit.
\label{fig:rv_folded}}
\end{figure*}

After analysing the best photometry and RV models, we carried out a global modelling of the data. To do this, we used a transit model that included two planets (whose phase-folded transits are depicted in Fig.\,\ref{fig:tr_folded}), a Keplerian model with two planets (whose phase-folded RV are displayed in Fig.\,\ref{fig:rv_folded}), a stellar activity model that used a GP with a dSHO kernel to model photometry (Fig.\,\ref{fig:LC_TESS}), and a QP kernel to model the RV datasets (Fig.\,\ref{fig:RV_CARM}). Although we would have preferred to use the same GP for the joint fit, the large number of data in TESS ($>$\,80000) forced us to use a kernel implemented in \texttt{celerite,}  which has a significantly lower computational cost. In the planetary models, the parameters of $T_c$ and $P$ share normal priors, while uniform priors were set for the rest of planetary parameters. For the activity models, we chose normal priors for the parameters of the covariance amplitudes, centred on the standard deviation of each dataset (consistent with the posteriors obtained in the independent analysis). The period of the process was shared by the two activity models, and the hyperparameters of $\eta_{L}$ and $\eta_{\omega}$ were shared between RV datasets. All the prior and posterior results of the joint fit and their derived parameters are shown in Table \ref{tab:joint-fit}.

\begin{table*}
\caption{Prior and posterior parameters for the joint fit.}
\label{tab:joint-fit}
\centering
\begin{tabular}{lccc}
\hline\hline
Parameter & Prior & Posterior(e=$\omega$=0) & Posterior (e,$\omega$ free) \\
\hline
$T_{c}^{b}$[BJD] & $\mathcal{N}$(2458845.373, 0.1) & 2458845.3728180$^{+0.0008229}_{-0.0008362}$ & 245845.3728382$^{+0.0008656}_{-0.0008565}$\\
$P^{b}$[d] & $\mathcal{N}$(7.108, 0.1) & 7.1079475$^{+0.0000089}_{-0.0000088}$ & 7.1079477$^{+0.0000087}_{-0.0000087}$\\
$R_{p}^{b}$[R$_{\mathrm{Jup}}$] & $\mathcal{U}$(0, 1) & 0.191$^{+0.008}_{-0.006}$ & 0.195$^{+0.008}_{-0.007}$\\
$b^{b}$ & $\mathcal{U}$(0, 1) & 0.266$^{+0.098}_{-0.141}$ & 0.342$^{+0.178}_{-0.214}$\\
$K^{b}$[m\,s$^{-1}$] & $\mathcal{U}$(0, 50) & $<$7.412 (3$\sigma$) & $<$7.574 (3$\sigma$)\\
$(\sqrt{e}\sin\omega)^{b}$ & $\mathcal{U}$(-1, 1) & -- & -0.092$^{+0.198}_{-0.281}$\\
$(\sqrt{e}\cos\omega)^{b}$ & $\mathcal{U}$(-1, 1) & -- & 0.002$^{+0.357}_{-0.351}$\\
\hline
$T_{c}^{c}$[BJD] & $\mathcal{N}$(2458844.058, 0.1) & 2458844.0589381$^{+0.0006400}_{-0.0006665}$ & 2458844.0589392$^{+0.0006803}_{-0.0007098}$\\
$P^{c}$[d] & $\mathcal{N}$(20.545, 0.1) & 20.5438281$^{+0.0000238}_{-0.0000236}$ & 20.5438295$^{+0.0000239}_{-0.0000240}$\\
$R_{p}^{c}$[R$_{\mathrm{Jup}}$] & $\mathcal{U}$(0, 1) & 0.240$^{+0.010}_{-0.008}$ & 0.242$^{+0.010}_{-0.009}$\\
$b^{c}$ & $\mathcal{U}$(0, 1) & 0.544$^{+0.041}_{-0.038}$ & 0.555$^{+0.058}_{-0.126}$\\
$K^{c}$[m\,s$^{-1}$] & $\mathcal{U}$(0, 50) & 3.737$^{+0.926}_{-0.913}$ & 3.780$^{+1.113}_{-0.991}$\\
$(\sqrt{e}\sin\omega)^{c}$ & $\mathcal{U}$(-1, 1) & -- & -0.013$^{+0.218}_{-0.217}$\\
$(\sqrt{e}\cos\omega)^{c}$ & $\mathcal{U}$(-1, 1) & -- & 0.001$^{+0.345}_{-0.329}$\\
\hline
$\gamma_{\mathrm{TESS}}$[ppt] & $\mathcal{U}$(-3$\sigma_{\mathrm{TESS}}$, 3$\sigma_{\mathrm{TESS}}$) & 0.131$^{+0.371}_{-0.374}$ & 0.144$^{+0.367}_{-0.368}$\\
$\sigma_{\mathrm{jit,TESS}}$[ppt] & $\mathcal{U}$(0, 3$\sigma_{\mathrm{TESS}}$) & 0.1153$^{+0.0008}_{-0.0008}$ & 0.1153$^{+0.0008}_{-0.0008}$\\
$\gamma_{\mathrm{CARMV}}$[m\,s$^{-1}$] & $\mathcal{U}$(-3$\sigma_{\mathrm{CARMV}}$, 3$\sigma_{\mathrm{\mathrm{CARMV}}}$) & 1.577$^{+3.905}_{-3.964}$ & 1.313$^{+4.036}_{-4.040}$\\
$\sigma_{\mathrm{jit,CARMV}}$[m\,s$^{-1}$] & $\mathcal{U}$(0, 3$\sigma_{\mathrm{CARMV}}$) & 0.658$^{+0.714}_{-0.462}$ & 0.645$^{+0.710}_{-0.450}$\\
$\gamma_{\mathrm{CARMN}}$[m\,s$^{-1}$] & $\mathcal{U}$(-3$\sigma_{\mathrm{CARMN}}$, 3$\sigma_{\mathrm{\mathrm{CARMN}}}$) & 5.751$^{+4.591}_{-4.638}$ & 5.730$^{+4.572}_{-4.604}$\\
$\sigma_{\mathrm{jit,CARMN}}$[m\,s$^{-1}$] & $\mathcal{U}$(0, 3$\sigma_{\mathrm{CARMN}}$) & 1.472$^{+1.634}_{-1.032}$ & 1.459$^{+1.619}_{-1.006}$\\
$\ln \eta_{\sigma_{1},\mathrm{TESS}}$ & $\mathcal{N}$($\ln\sigma_{\mathrm{TESS}}$, 0.5) & 1.370$^{+0.087}_{-0.075}$ & 1.368$^{+0.091}_{-0.076}$\\
$\ln \eta_{\sigma_{2},\mathrm{TESS}}$ & $\mathcal{N}$($\ln\sigma_{\mathrm{TESS}}$, 0.5) & 1.106$^{+0.393}_{-0.334}$ & 1.114$^{+0.389}_{-0.334}$\\
$\ln \eta_{L_{1},\mathrm{TESS}}$ & $\mathcal{U}$($\ln P_{\mathrm{rot}}/2\pi$, 10) & 1.227$^{+0.191}_{-0.161}$ & 1.226$^{+0.197}_{-0.162}$\\
$\ln \eta_{L_{2},\mathrm{TESS}}$ & $\mathcal{U}$($\ln P_{\mathrm{rot}}/2\pi$, 10) & 7.100$^{+1.496}_{-1.407}$ & 7.037$^{+1.556}_{-1.403}$\\
$\ln \eta_{P_{\mathrm{rot}}}$ & $\mathcal{U}$(0, 2.3) & 1.857$^{+0.001}_{-0.003}$ & 1.857$^{+0.001}_{-0.003}$\\
$\ln \eta_{\sigma, \mathrm{CARMV}}$ & $\mathcal{N}$($\ln\sigma_{\mathrm{CARMV}}$, 0.5) & 3.046$^{+0.102}_{-0.096}$ & 3.054$^{+0.102}_{-0.095}$\\
$\ln \eta_{\sigma, \mathrm{CARMN}}$ & $\mathcal{N}$($\ln\sigma_{\mathrm{CARMN}}$, 0.5) & 3.117$^{+0.108}_{-0.105}$ & 3.122$^{+0.106}_{-0.107}$\\
$\ln \eta_{L, \mathrm{RV}}$ & $\mathcal{U}$($\ln P_{\mathrm{rot}}/2\pi$, 10) & 2.973$^{+0.075}_{-0.077}$ & 2.961$^{+0.072}_{-0.075}$\\
$\eta_{\omega, \mathrm{RV}}$ & $\mathcal{U}$(0.1, 2.0) & 0.253$^{+0.022}_{-0.022}$ & 0.254$^{+0.022}_{-0.021}$\\
$q_{1,\mathrm{TESS}}$ & $\mathcal{N}$(0.3, 0.1) & 0.228$^{+0.056}_{-0.049}$ & 0.221$^{+0.056}_{-0.048}$\\
$q_{2,\mathrm{TESS}}$ & $\mathcal{N}$(0.37, 0.1) & 0.318$^{+0.091}_{-0.089}$ & 0.331$^{+0.096}_{-0.096}$\\
$M_{\star}$[M$_{\odot}$] & $\mathcal{N}$(0.956, 0.022) & 0.957$^{+0.022}_{-0.022}$ & 0.956$^{+0.022}_{-0.022}$\\
$R_{\star}$[R$_{\odot}$] & $\mathcal{N}$(0.934, 0.029) & 0.929$^{+0.030}_{-0.024}$ & 0.939$^{+0.029}_{-0.028}$\\
\hline
\multicolumn{4}{c}{Derived parameters}\\
\hline
$R_{p}^{b}/R_{\star}$ & -- & 0.0207$^{+0.0003}_{-0.0003}$ & 0.0208$^{+0.0005}_{-0.0004}$\\
$R_{p}^{b}$ [R$_{\oplus}$] & -- & 2.140$^{+0.087}_{-0.069}$ & 2.183$^{+0.089}_{-0.082}$\\
$a^{b}/R_{\star}$ & -- & 16.500$^{+0.459}_{-0.541}$ & 16.319$^{+0.517}_{-0.508}$\\
$i^{b}$ [$^\circ$] & -- & 89.075$^{+0.502}_{-0.381}$ & 88.800$^{+0.756}_{-0.629}$\\
$e^{b}$ & -- & -- & 0.129$^{+0.170}_{-0.094}$\\
$\omega^{b}$ [rad] & -- & -- & -0.646$^{+2.374}_{-1.878}$\\
$M_p^{b}$ [M$_{\oplus}$] & -- & $<$21.755(3$\sigma$) & $<$21.888(3$\sigma$)\\
$\rho^{b}$ [g\,cm$^{-3}$] & -- & $<$12.998(3$\sigma$) & $<$13.484(3$\sigma$)\\
$T_{eq}^{b}(A=0)$ [K] & -- & 967.11$^{+18.50}_{-15.89}$ & 972.014$^{+18.331}_{-17.877}$\\
$T_{eq}^{b}(A=0.6)$ [K] & -- & 769.12$^{+14.72}_{-12.64}$ & 773.01$^{+14.58}_{-14.22}$\\
\hline
$R_{p}^{c}/R_{\star}$ & -- & 0.0260$^{+0.0004}_{-0.0004}$ & 0.0260$^{+0.0005}_{-0.0005}$\\
$R_{p}^{c}$ [R$_{\oplus}$] & -- & 2.692$^{+0.108}_{-0.088}$ & 2.718$^{+0.108}_{-0.101}$\\
$a^{c}/R_{\star}$ & -- & 33.479$^{+0.931}_{-1.098}$ & 33.113$^{+1.049}_{-1.031}$\\
$i^{c}$ [$^\circ$] & -- & 89.070$^{+0.089}_{-0.104}$ & 89.045$^{+0.226}_{-0.130}$\\
$e^{c}$ & -- & -- & 0.097$^{+0.284}_{-0.075}$\\
$\omega^{c}$ [rad] & -- & -- & -0.098$^{+2.364}_{-2.265}$\\
$M_p^{c}$ [M$_{\oplus}$] & -- & 15.539$^{+3.861}_{-3.802}$ & 15.319$^{+4.188}_{-3.992}$\\
$\rho^{c}$ [g\,cm$^{-3}$] & -- & 4.630$^{+1.297}_{-1.200}$ & 4.453$^{+1.414}_{-1.220}$\\
$T_{eq}^{c}(A=0)$ [K] & -- & 678.92$^{+13.04}_{-11.11}$ & 682.37$^{+12.87}_{-12.58}$\\
$T_{eq}^{c}(A=0.6)$ [K] & -- & 539.92$^{+10.37}_{-8.85}$ & 542.67$^{+10.24}_{-10.01}$\\
\hline
$u_{1,\mathrm{TESS}}$ & -- & 0.301$^{+0.074}_{-0.076}$ & 0.307$^{+0.079}_{-0.077}$\\
$u_{2,\mathrm{TESS}}$ & -- & 0.173$^{+0.103}_{-0.092}$ & 0.157$^{+0.108}_{-0.093}$\\
\hline\hline

\hline
\multicolumn{3}{l}{\small{The prior label of $\mathcal{N}$ and $\mathcal{U}$ represent normal and uniform distribution, respectively.}}\\
\multicolumn{3}{l}{\small{"CARMV" refers to CARMENES VIS and "CARMN" refers to CARMENES NIR.}} \\
\end{tabular}
\end{table*}

\section{Discussion}
\label{sec:disc}

\subsection{Densities and compositions}

From the posterior parameters obtained in the joint fit, we derive an upper limit of $M_p^b<$\,21.8 M$_\oplus$ on the mass of HD\,63433\,b at the level of 3$\sigma$ and present a 4$\sigma$ detection for HD\,63433\,c, yielding a mass of $M_p^c$=\,15.5\,$\pm$\,3.9 M$_\oplus$. These values, together with the radii of planets b and c updated in this work, give bulk densities of $\rho^{b}$\,$<$\,13.0 g\,cm$^{-3}$ and $\rho^{c}$\,=\,4.6\,$\pm$\,1.3 g\,cm$^{-3}$, respectively. The equilibrium temperature ($T_{\rm{eq}}$) for both planets lies in the range of 769--967\,K for HD\,63433\,b and 540--679\,K for HD\,63433\,c, assuming planetary albedos (A) between 0 and 0.6\@. Our results imply that both planets are warm sub-Neptunes. The density of HD\,63433\,c is slightly lower than that of Earth. 

\begin{figure*}[ht!]
\includegraphics[width=1\linewidth]{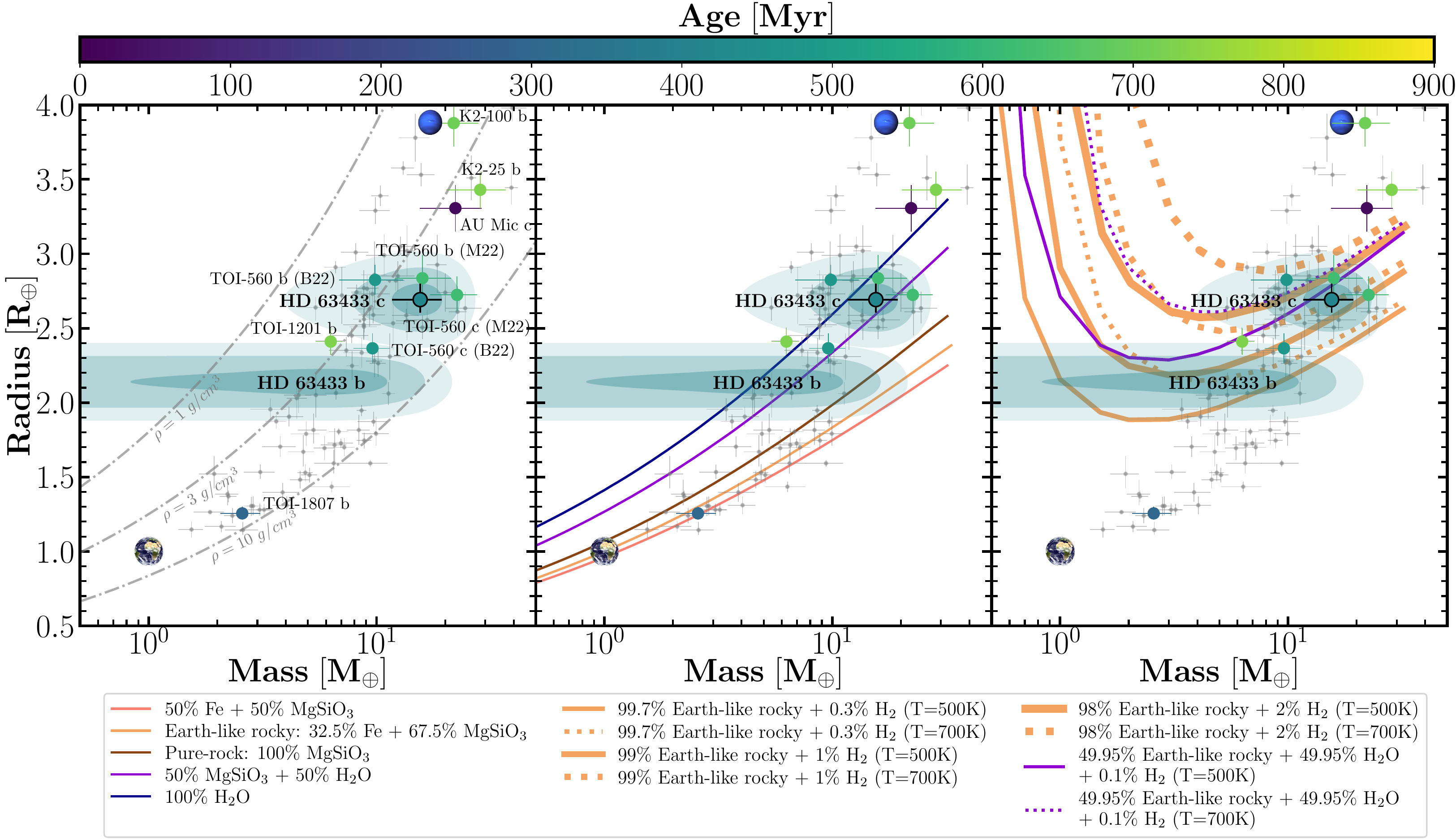}
\caption{Radius-mass diagram of HD\,63433\,b and c, together with all exoplanets (grey dots) with a precision better than 8\% in radius (by transits) and 20\% in mass (by RV). Young exoplanets with measured masses are plotted as coloured dots according to their ages. The uncertainties on the HD\,63433\,b and c planets are shown as coloured shaded regions with 1, 2, and 3$\sigma$ levels of confidence. The dashed grey lines are constant density lines. Coloured lines indicate different composition models without gas (middle panel) and with a gas envelope (right panel) from \cite{Zeng19}. Earth and Neptune are also depicted as reference. B22 and M22 refer to \cite{barr22} and \cite{elmufti22}, respectively.
\label{fig:RM}}
\end{figure*}

Fig.\,\ref{fig:RM} shows a mass-radius diagram that presents all known small exoplanets (taken from the Extrasolar Planets Encyclopaedia\footnote{http://exoplanet.eu/}) with radius precisions better than 8\% (measured through the transit method) and mass precisions better than 20\% (measured through RV). The uncertainties of HD\,63433\,b and c are represented as a coloured-shaded region, according to the 1, 2, and 3$\sigma$ levels of confidence. To place the planets in context, we also plot the young planets ($\le$900 Myr) as coloured dots according to their age. Only eight young planets have a measured radius and mass (AU\,Mic\,c, TOI-1807\,b, TOI-560\,b and c, K2-25\,b, K2-100\,b, TOI-1201\,b, and now we add HD\,63433\,c). They represent less than 10\% of the total population of exoplanet shown in the diagram. The left panel of Fig.\ \ref{fig:RM} shows that the planets are not randomly distributed. Most of them are concentrated between the density lines of 1--10 g\,cm$^{-3}$ with two main groups: those with masses lower than 8--10 M$_\oplus$ with densities of 3--10 g\,cm$^{-3}$, and those with higher masses, with typically a larger radius and lower densities. Young planets also follow this distribution so far, although only one planet, TOI-1807\,b, currently has the characteristic radius and mass of super-Earths. HD\,63433\,c also matches the planets in the field. Its density is only slightly lower than that of Earth, but is consistent within the error bars.

Radius and mass measurements can help to determine the internal composition of the planets. We plot models of different compositions by mass without an atmosphere (in the middle panel) and with atmospheres (on the right) from \cite{Zeng19} in Fig.\,\ref{fig:RM}. In the first case, where the internal composition of the planets does not have a significant envelope, our measurements allow us to rule out that HD\,63433\,b contains more than 50\% of iron by mass, and it may be compatible with planets with a higher silicate or water content. However, the composition of HD\,63433\,c does not contain a significant mass fraction of iron and is most probably formed by silicates and/or water. On the other hand, models with envelopes can only explain their positions in the mass-radius diagram if their atmospheres contain 2\% of H$_2$ by mass at most for any composition mixture.  

\subsection{Formation history}

Formation models of planets with orbital periods shorter than 100 days suggest that the most massive planets (with a core mass greater than 5--10 M$_\oplus$) should form in outer regions (beyond the ice line) where the mean size of ice pebbles is greater than the size of the silicates \citep{morbidelli15, venturini20}. Higher masses in this region also favour gas accretion, which notably increases the size of the planets. Then, interactions with the disc gas are expected to trigger the migration of several of these planets into the inner regions, causing them to mix with rocky planets that have already formed there. This mechanism is expected to occur on timescales shorter than 10 Myr, which is the typical lifetime of protoplanetary mass discs \citep{williams11}. The discovery of a Neptune-size planet in the Upper Scorpius region (5--10\,Myr) confirms this \citep{david16,mann16b}. At this stage, evolutionary mechanisms such as photoevaporation \citep{owen17}, which takes place on timescales of $\text{several }$tens of million years, or core-powered mass-loss \citep{ginzburg18,gupta20}, which acts on timescales of $\sim$\,1 Gyr, may play an important role in removing part of these primordial atmospheres of short-period planets.

Young, multi-planet systems offer a unique opportunity to study the formation and the effects of evolution on planets. HD\,63433 has an estimated intermediate age of $\sim$400 Myr and therefore constitutes a good example on which these mechanisms can be tested. The mass measurement of HD\,63433\,c and the upper limit set on the mass of HD\,63433\,b presented in this paper allow us to determine the composition of these planets and discuss them in the context of their evolution. In this sense, planet HD\,63433\,c is mostly composed of rocks and water, with little or no gas envelope, while HD\,63433\,b may have a similar composition or a larger part of gas. Recently, \cite{zhang22} have shown that the atmosphere of both planets does not show signs of an intense evaporation. These results suggest that if HD\,63433\,c had a significant gas envelope in the past, as expected for such a massive planet according to formation models, it has already lost most of its gas content at the age of system. This favours rapid evolutionary mechanisms, such as photo-evaporation, which can explain the mass loss of the planet atmospheres on timescales shorter than a few hundred million years.

All the young planets shown in Fig.\,\ref{fig:RM}, except for TOI-1807 (catalogued as an ultra-short period planet), are on the upper side of the radius valley (R\,$>$\,1.9 R$_\oplus$), which corresponds to Neptune/sub-Neptune planets whose composition could be similar to HD\,63433\,c. This is probably an observational bias, as mentioned earlier, because of the high stellar activity of young stars that makes the detection and measurements of the physical properties of small planets around these stars difficult. Therefore, young planets are typically found with larger radii and higher masses because they are easier to detect. A recent work \citep{luque22} has shown that the population of small planets orbiting M dwarfs is composed of three types of planets with different compositions: rocky planets, water worlds (50\% water, 50\% silicates), and small gas planets. Although HD\,63433\,c (and possibly HD\,63433\,b also) and other young small planets, such as TOI-1201\,b, TOI-560\,b and c, seem to follow the sequence of water planets, while more massive young planets, such as K2-100\,b or AU\,Mic\,b and c, seem to be small gas planets, more similar to Neptune and Uranus, the number is still low. It is necessary to obtain more dynamical mass determinations of young planets to compare the atmospheric compositions of young and old populations.

\subsection{Dependence on flux and RV}

\begin{figure*}[ht!]
\includegraphics[width=1\linewidth]{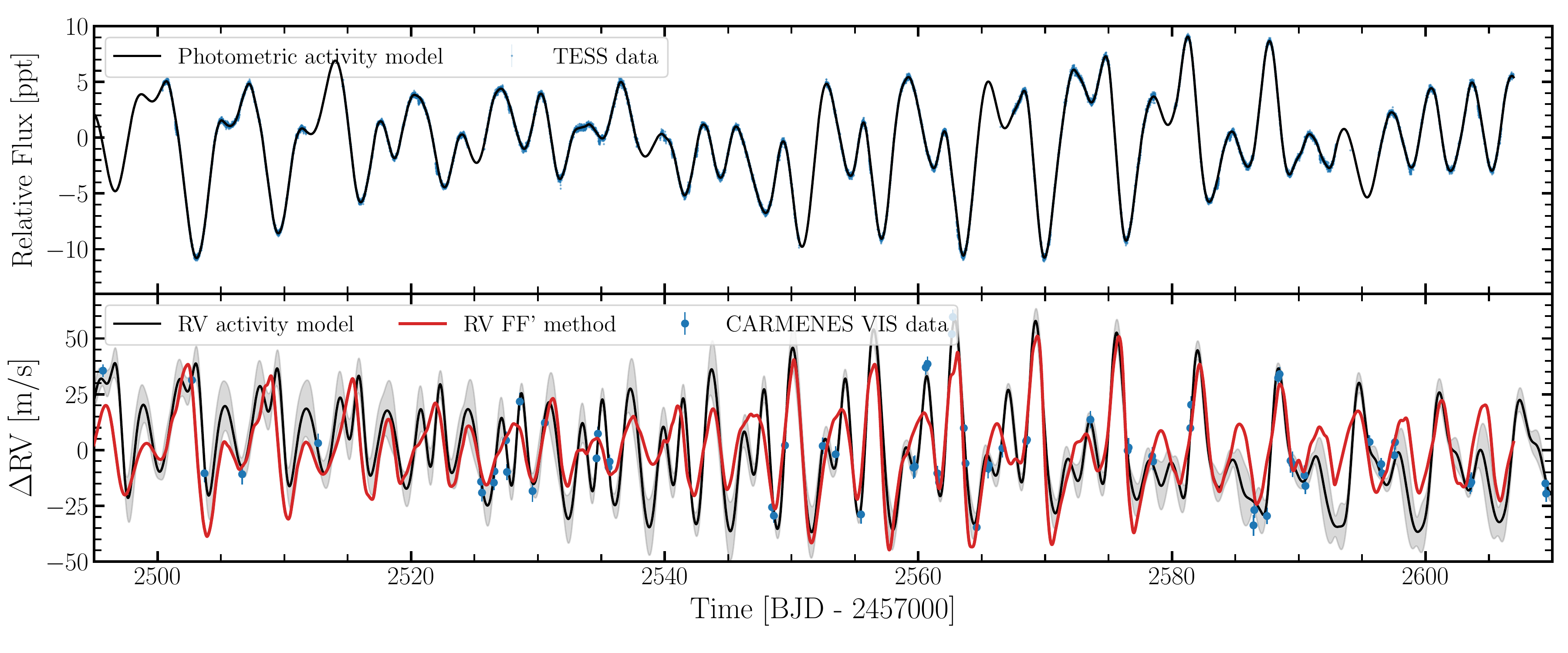}
\caption{Comparison between photometric and RV stellar activity. \textit{Top panel}: TESS light curve of HD\,63433 contemporaneously with the RV of CARMENES data. The blue dots are the PDCSAP flux, and the black line indicates the best activity model. \textit{Bottom panel}: CARMENES RV VIS data shown with blue dots. The black line shows the GP activity model with its 1$\sigma$ confidence level (grey shadow). The solid red line depicts the RV model predicted from the TESS light curve using FF$^\prime$ method..
\label{fig:FF}}
\end{figure*}

\cite{aigrain12} presented a method with analytical formulas to predict the stellar activity in RV from the observed photometric fluxes. The high activity of young stars makes them an ideal laboratory for testing this FF$^\prime$ method. Because part of our observations with CARMENES were intentionally obtained simultaneously with TESS, we can observationally verify the behaviour of stellar activity in photometry and RV. The upper panel of Fig.\ \ref{fig:FF} shows the TESS data contemporaneously with the CARMENES RVs and the activity model obtained in the transit-only fit. The lower panel of Fig.\ \ref{fig:FF} displays the CARMENES VIS data together with the activity model (solid black line) obtained in the RV-only fit for the same time interval as in the upper panel. The stellar activity in TESS presents a double modulation, in the same way as the RV. This agrees with the analysis of the RV periodograms, whose the maximum peak is at 3.2d, in agreement with the first harmonic of the rotation period. In addition, in the RV observations where the cadence is better (between 2459550 and 2459570 BJD), the model presents up to three peaks per rotation period. This also agrees with the analysis of the periodograms of the RV, and some indicators also show a significant peak at 2.1 days. Using the analytical formulas in \cite{aigrain12} based on a simple spot model, we calculated the $\Delta$RV$_{\rm{rot}}$ from the stellar activity modeled in TESS. In the same panel, we represent the result of the FF$^\prime$ method (normalized by a scale factor) with a solid red line together with the observed RV data. Both RV models, the one calculated from GPs and the other calculated with the FF$^\prime$ method from photometry, behave similarly throughout the time series. Although the FF$^\prime$ model never reproduces three peaks per rotation period, both models reproduce the amplitude variations between one period and another and between the two main peaks within each period in the same way. That is, the FF$^\prime$ method predicts  similar stellar activity as the GP model based on photometry and with the $\Delta$RV$_{\rm{rot}}$ term alone, suggesting that most of the stellar activity affecting the photometry also affects the RV in a similar way. Therefore, obtaining contemporaneous photometry with high cadence as in TESS together with RV data maybe be a valuable resource for characterizing the source of stellar activity and correct it with the appropriate models to determine the masses of young or low-mass planets.

\section{Conclusions}
\label{sec:concl}

We have analysed precise photometric light curves of five sectors of the TESS mission and obtained $\sim$150 precise RV measurements with CARMENES VIS and NIR of the HD\,63433 star. We performed a joint-fit analysis of these data and present an update of the transit parameters and a characterization of the dynamic masses of the HD\,63433 planetary system. The inner planet, HD\,63433\,b, has an orbital period of P$^b$\,=\,7.108 days, a radius of R$_p^b$\,=\,2.140\,$\pm$\,0.087 R$_\oplus$, and an upper limit at 3$\sigma$ level for the amplitude of K$_p^b$\,$<$\,7.41 m\,s$^{-1}$. From these values, we derive a mass of $M_p^b$\,$<$\,21.76 M$_\oplus$ and a planetary density of $\rho^b$\,$<$\,13.00 $\rho_\oplus$. On the other hand, for planet HD\,63433\,c, we obtain a period of P$^c$\,=\,20.544 days, with a radius of R$_p^c$\,=\,2.692\,$\pm$\,0.108 R$_\oplus$ and an amplitude of K$_p^c$\,=\,3.74\,$\pm$\,0.93 m\,s$^{-1}$, translating into a mass of $M_p^c$\,=\,15.54\,$\pm$\,3.86 M$_\oplus$ and a density $\rho^c$\,=\,4.63\,$\pm$\,1.30 $\rho_\oplus$.

According to theoretical models, planet HD\,63433\,c is composed mostly of rocks and water and  has little or no gas envelope, while HD\,63433\,b may have a similar composition, but its gas content remains unconstrained due to the lack of a mass measurement. These results, together with the lack of signs of intense evaporation in both planets (\cite{zhang22}), suggest that if HD\,63433\,c had a significant gas envelope in the past, it has already lost most of it at about 400 Myr. This favours rapid evolutionary mechanism of mass loss, such as photoevaporation.

We acknowledge that HD\,63433 has been followed up also with the HARPS-N spectrograph \citep{damasso23}. Our team and the GAPS team have coordinated the submission of two studies, which were intentionally carried out in an independent way.

\begin{acknowledgements}
MMD, NL, VJSB, MRZO, HT, ASM, EP, and DM acknowledge  support  from  the Agencia  Estatal  de  Investigaci\'on  del  Ministerio  de  Ciencia  e Innovaci\'on (AEI-MCINN) under grant PID2019-109522GB-C5[1,3,4]\@. 
CARMENES is an instrument for the Centro Astron\'{o}mico Hispano-Alem\'{a}n de Calar Alto (CAHA, Almer\'{\i}a, Spain). CARMENES is funded by the German Max-Planck-Gesellschaft (MPG), the Spanish Consejo Superior de Investigaciones Cient\'{\i}ficas (CSIC), the European Union through FEDER/ERF FICTS-2011-02 funds, and the members of the CARMENES Consortium (Max-Planck-Institut f\"{u}r Astronomie, Instituto de Astrof\'{\i}sica de Andaluc\'{\i}a, Landessternwarte K\"{o}nigstuhl, Institut de Ci\`{e}ncies de l'Espai, Institut f\"{u}r Astrophysik G\"{o}ttingen, Universidad Complutense de Madrid, Th\"{u}ringer Landessternwarte Tautenburg, Instituto de Astrof\'{\i}sica de Canarias, Hamburger Sternwarte, Centro de Astrobiolog\'{\i}a and Centro Astron\'{o}mico Hispano-Alem\'{a}n), with additional contributions by the Spanish Ministry of Economy, the German Science Foundation through the Major Research Instrumentation Programme and DFG Research Unit FOR2544 "Blue Planets around Red Stars", the Klaus Tschira Stiftung, the states of Baden-W\"{u}rttemberg and Niedersachsen, and by the Junta de Andaluc\'{\i}a.
Based on observations collected at the Centro Astronómico Hispano-Alemán (CAHA) at Calar Alto, operated jointly by Junta de Andalucía and Consejo Superior de Investigaciones Científicas (IAA-CSIC) under programmes H20-3.5-020 and F21-3.5-010\@.
Part of this work was supported by the German \emph{Deut\-sche For\-schungs\-ge\-mein\-schaft, DFG\/} project number Ts~17/2--1\@.
We acknowledge the use of public TESS data from pipelines at the TESS Science Office and at the TESS Science Processing Operations Center. 
Resources supporting this work were provided by the NASA High-End Computing (HEC) Program through the NASA Advanced Supercomputing (NAS) Division at Ames Research Center for the production of the SPOC data products.
This paper includes data collected with the TESS mission, obtained from the MAST data archive at the Space Telescope Science Institute (STScI). Funding for the TESS mission is provided by the NASA Explorer Program. STScI is operated by the Association of Universities for Research in Astronomy, Inc., under NASA contract NAS 5–26555.
This research has made use of the Simbad and Vizier databases, operated at the centre de Donn\'ees Astronomiques de Strasbourg (CDS), and of NASA's Astrophysics Data System Bibliographic Services (ADS).

\end{acknowledgements}

\bibliographystyle{aa.bst} 
\bibliography{biblio.bib}

\begin{appendix}

\section{Transit-only fit}
\label{appendixA}

\begin{table*}
\caption{Prior and posterior parameters for the transit-only fit.}
\label{tab:tronly-fit}
\centering
\begin{tabular}{lccc}
\hline\hline
Parameter & Prior & Posterior(e=$\omega$=0) & Posterior (e,$\omega$ free) \\
\hline
$T_{c}^{b}$[BJD] & $\mathcal{N}$(2458845.373, 0.1) & 2458845.37282$^{+0.00083}_{-0.00083}$ & 2458845.37281$^{+0.00098}_{-0.00103}$\\
$P^{b}$[d] & $\mathcal{N}$(7.108, 0.1) & 7.1079474$^{+0.0000089}_{-0.0000088}$ & 7.1079475$^{+0.0000088}_{-0.0000086}$\\
$R_{p}^{b}$[R$_{\mathrm{Jup}}$] & $\mathcal{U}$(0, 1) & 0.191$^{+0.008}_{-0.006}$ & 0.195$^{+0.009}_{-0.007}$\\
$b^{b}$ & $\mathcal{U}$(0, 1) & 0.265$^{+0.098}_{-0.142}$ & 0.38652$^{+0.182}_{-0.242}$\\
$(\sqrt{e}\sin\omega)^{b}$ & $\mathcal{U}$(-1, 1) & -- & -0.178$^{+0.248}_{-0.293}$\\
$(\sqrt{e}\cos\omega)^{b}$ & $\mathcal{U}$(-1, 1) & -- & -0.007$^{+0.464}_{-0.490}$\\
\hline
$T_{c}^{c}$[BJD] & $N$ (2458844.058, 0.1) & 2458844.05895$^{+0.00065}_{-0.00066}$ & 2458844.05891$^{+0.00080}_{-0.00081}$\\
$P^{c}$[d] & $\mathcal{N}$(20.545, 0.1) & 20.5438280$^{+0.0000239}_{-0.0000239}$ & 20.5438297$^{+0.0000251}_{-0.0000246}$\\
$R_{p}^{c}$[R$_{\mathrm{Jup}}$] & $\mathcal{U}$(0, 1) & 0.240$^{+0.010}_{-0.008}$ & 0.242$^{+0.010}_{-0.010}$\\
$b^{c}$ & $\mathcal{U}$(0, 1) & 0.543$^{+0.041}_{-0.038}$ & 0.567$^{+0.081}_{-0.163}$\\
$(\sqrt{e}\sin\omega)^{c}$ & $\mathcal{U}$(-1, 1) & -- & -0.074$^{+0.285}_{-0.252}$\\
$(\sqrt{e}\cos\omega)^{c}$ & $\mathcal{U}$(-1, 1) & -- & -0.023$^{+0.558}_{-0.497}$\\
\hline
$\gamma_{\mathrm{TESS}}$[ppt] & $\mathcal{U}$(-3$\sigma_{\mathrm{TESS}}$, 3$\sigma_{\mathrm{TESS}}$) & 0.125$^{+0.369}_{-0.368}$ & 0.130$^{+0.363}_{-0.361}$\\
$\sigma_{\mathrm{jit,TESS}}$[ppt] & $\mathcal{U}$(0, 3$\sigma_{\mathrm{TESS}}$) & 0.1153$^{+0.0008}_{-0.0008}$ & 0.1153$^{+0.0008}_{-0.0008}$\\
$\ln \eta_{\sigma_{1},\mathrm{TESS}}$ & $\mathcal{N}$($\ln\sigma_{\mathrm{TESS}}$, 0.5) & 1.371$^{+0.089}_{-0.078}$ & 1.369$^{+0.087}_{-0.078}$\\
$\ln \eta_{\sigma_{2},\mathrm{TESS}}$ & $\mathcal{N}$($\ln\sigma_{\mathrm{TESS}}$, 0.5) & 1.140$^{+0.394}_{-0.332}$ & 1.129$^{+0.412}_{-0.338}$\\
$\ln \eta_{L_{1},\mathrm{TESS}}$ & $\mathcal{U}$($\ln P_{\mathrm{rot}}/2\pi$, 10) & 1.240$^{+0.203}_{-0.168}$ & 1.233$^{+0.203}_{-0.169}$\\
$\ln \eta_{L_{1},\mathrm{TESS}}$ & $\mathcal{U}$($\ln P_{\mathrm{rot}}/2\pi$, 10) & 7.257$^{+1.582}_{-1.643}$ & 7.262$^{+1.593}_{-1.665}$\\
$\ln P_{\mathrm{rot}}$ & $\mathcal{U}$(0, 2.3) & 1.854$^{+0.003}_{-0.002}$ & 1.854$^{+0.004}_{-0.002}$\\
$q_{1,\mathrm{TESS}}$ & $\mathcal{N}$(0.3, 0.1) & 0.227$^{+0.056}_{-0.049}$ & 0.221$^{+0.056}_{-0.048}$\\
$q_{2,\mathrm{TESS}}$ & $\mathcal{N}$(0.37, 0.1) & 0.317$^{+0.092}_{-0.090}$ & 0.332$^{+0.101}_{-0.094}$\\
$M_{\star}$[M$_{\odot}$] & $\mathcal{N}$(0.956, 0.022) & 0.957$^{+0.022}_{-0.022}$ & 0.955$^{+0.022}_{-0.021}$\\
$R_{\star}$[R$_{\odot}$] & $\mathcal{N}$(0.934, 0.029) & 0.929$^{+0.030}_{-0.024}$ & 0.937$^{+0.029}_{-0.029}$\\
\hline
\multicolumn{4}{c}{Derived parameters}\\
\hline
$R_{p}^{b}/R_{\star}$ & -- & 0.0207$^{+0.0003}_{-0.0003}$ & 0.0209$^{+0.0006}_{-0.0004}$\\
$R_{p}^{b}$ [R$_{\oplus}$] & -- & 2.139$^{+0.086}_{-0.069}$ & 2.186$^{+0.096}_{-0.084}$\\
$a^{b}/R_{\star}$ & -- & 16.506$^{+0.458}_{-0.541}$ & 16.345$^{+0.530}_{-0.500}$\\
$i^{b}$ [$^\circ$] & -- & 89.081$^{+0.505}_{-0.383}$ & 88.641$^{+0.856}_{-0.660}$\\
$e^{b}$ & -- & -- & 0.212$^{+0.301}_{-0.166}$\\
$\omega^{b}$ [rad] & -- & -- & -0.850$^{+1.805}_{-1.757}$\\
$R_{p}^{c}/R_{\star}$ & -- & 0.0260$^{+0.0004}_{-0.0004}$ & 0.0260$^{+0.0006}_{-0.0005}$\\
$R_{p}^{c}$ [R$_{\oplus}$] & -- & 2.691$^{+0.109}_{-0.088}$ & 2.716$^{+0.107}_{-0.108}$\\
$a^{c}/R_{\star}$ & -- & 33.490$^{+0.928}_{-1.098}$ & 33.164$^{+1.075}_{-1.014}$\\
$i^{c}$ [$^\circ$] & -- & 89.071$^{+0.088}_{-0.104}$ & 89.025$^{+0.286}_{-0.170}$\\
$e^{c}$ & -- & -- & 0.200$^{+0.314}_{-0.138}$\\
$\omega^{c}$ [rad] & -- & -- & -0.320$^{+2.463}_{-2.329}$\\
\hline
$u_{1,\mathrm{TESS}}$ & -- & 0.299$^{+0.075}_{-0.076}$ & 0.310$^{+0.079}_{-0.080}$\\
$u_{2,\mathrm{TESS}}$ & -- & 0.173$^{+0.105}_{-0.092}$ & 0.156$^{+0.106}_{-0.098}$\\
\hline\hline

\hline
\multicolumn{4}{l}{\small{The prior label of $\mathcal{N}$ and $\mathcal{U}$ represent normal and uniform distribution, respectively.}}\\
\end{tabular}
\end{table*}

\section{RV-only fit}
\label{appendixB}

\begin{table*}
\caption{Prior and posterior parameters for the RV-only fit.}
\label{tab:rvonly-fit}
\centering
\begin{tabular}{lccc}
\hline\hline
Parameter & Prior & Posterior(e=$\omega$=0) & Posterior (e,$\omega$ free) \\
\hline
$T_{c}^{b}$[BJD] & $\mathcal{N}$(2458845.373, 0.1) & 2458845.3742760$^{+0.0995880}_{-0.1004927}$ & 2458845.3669810$^{+0.0991998}_{-0.1022971}$\\
$P^{b}$[d] & $\mathcal{N}$(7.108, 0.01) & 7.1072690$^{+0.0080481}_{-0.0077893}$ & 7.1015434$^{+0.0125120}_{-0.0062255}$\\
$K^{b}$[m\,s$^{-1}$] & $\mathcal{U}$(0, 50) & $<$7.252 (3$\sigma$) & $<$15.430 (3$\sigma$)\\
$(\sqrt{e}\sin\omega)^{b}$ & $\mathcal{U}$(-1, 1) & -- & 0.020$^{+0.611}_{-0.605}$\\
$(\sqrt{e}\cos\omega)^{b}$ & $\mathcal{U}$(-1, 1) & -- & -0.031$^{+0.606}_{-0.564}$\\
$T_{c}^{c}$[BJD] & $\mathcal{N}$(2458844.058, 0.1) & 2458844.0567302$^{+0.0987396}_{-0.0994477}$ & 2458844.0613962$^{+0.0978948}_{-0.1011432}$\\
$P^{c}$[d] & $\mathcal{N}$(20.545, 0.1) & 20.5301384$^{+0.0326166}_{-0.0297178}$ & 20.5170526$^{+0.0321096}_{-0.0258296}$\\
$K^{c}$[m\,s$^{-1}$] & $\mathcal{U}$(0, 50) & 3.687$^{+0.977}_{-0.960}$ & 4.439$^{+3.606}_{-1.422}$\\ 
$(\sqrt{e}\sin\omega)^{c}$ & $\mathcal{U}$(-1, 1) & -- & 0.055$^{+0.525}_{-0.480}$\\
$(\sqrt{e}\cos\omega)^{c}$ & $\mathcal{U}$(-1, 1) & -- & 0.010$^{+0.500}_{-0.488}$\\
\hline
$\gamma_{\mathrm{CARMV}}$[m\,s$^{-1}$] & $\mathcal{U}$(-3$\sigma_{\mathrm{CARMV}}$, 3$\sigma_{\mathrm{CARMV}}$) & 1.605$^{+3.952}_{-4.031}$ & 1.248$^{+4.094}_{-4.131}$\\
$\sigma_{\mathrm{jit,CARMV}}$[m\,s$^{-1}$] & $\mathcal{U}$(0, 3$\sigma_{\mathrm{CARMV}}$) & 0.637$^{+0.701}_{-0.448}$ & 0.637$^{+0.688}_{-0.451}$\\
$\gamma_{\mathrm{CARMN}}$[m\,s$^{-1}$] & $\mathcal{U}$(-3$\sigma_{\mathrm{CARMN}}$, 3$\sigma_{\mathrm{CARMN}}$) & 5.817$^{+4.558}_{-4.595}$ & 5.780$^{+4.653}_{-4.706}$\\
$\sigma_{\mathrm{jit,CARMN}}$[m\,s$^{-1}$] & $\mathcal{U}$(0, 3$\sigma_{\mathrm{CARMN}}$) & 1.462$^{+1.621}_{-1.026}$ & 1.457$^{+1.610}_{-1.027}$\\
$\ln \eta_{\sigma, \mathrm{CARMV}}$ & $\mathcal{N}$($\ln\sigma_{\mathrm{CARMV}}$, 0.5) & 3.052$^{+0.102}_{-0.094}$ & 3.066$^{+0.106}_{-0.096}$\\
$\ln \eta_{\sigma, \mathrm{CARMN}}$ & $\mathcal{N}$($\ln\sigma_{\mathrm{CARMN}}$, 0.5) & 3.120$^{+0.107}_{-0.105}$ & 3.126$^{+0.110}_{-0.103}$\\
$\ln \eta_{L, \mathrm{RV}}$ & $\mathcal{U}$($\ln P_{\mathrm{rot}}/2\pi$, 10) & 2.954$^{+0.075}_{-0.077}$ & 2.957$^{+0.083}_{-0.081}$\\
$\eta_{\omega, \mathrm{RV}}$ & $\mathcal{U}$(0.1, 2.0) & 0.254$^{+0.022}_{-0.021}$ & 0.257$^{+0.022}_{-0.021}$\\
$\ln P_{\mathrm{rot}}$ & $\mathcal{U}$(0, 2.3) & 1.858$^{+0.003}_{-0.003}$ & 1.858$^{+0.004}_{-0.003}$\\
\hline
\multicolumn{4}{c}{Derived parameters}\\
\hline
$a^{b}/R_{\star}$ & -- & 16.401$^{+0.543}_{-0.508}$ & 16.401$^{+0.539}_{-0.511}$\\
$M_p\sin(i)^{b}$ [M$_{\oplus}$] & -- & $<$21.131 (3$\sigma$) & $<$22.460 (3$\sigma$)\\
$e^{b}$ & -- & -- & 0.541$^{+0.346}_{-0.329}$\\
$\omega^{b}$ [rad] & -- & -- & 0.067$^{+2.076}_{-2.228}$\\
$a^{c}/R_{\star}$ & -- & 33.266$^{+1.104}_{-1.030}$ & 33.263$^{+1.095}_{-1.037}$\\
$M_p\sin(i)^{c}$ [M$_{\oplus}$] & -- & 15.304$^{+4.077}_{-3.979}$ & 16.477$^{+5.912}_{-4.879}$\\ 
$e^{c}$ & -- & -- & 0.421$^{+0.369}_{-0.376}$\\
$\omega^{c}$ [rad] & -- & -- & 0.363$^{+1.892}_{-2.277}$\\
\hline\hline

\hline
\multicolumn{4}{l}{\small{The prior label of $\mathcal{N}$ and $\mathcal{U}$ represent normal and uniform distribution, respectively.}}\\
\multicolumn{4}{l}{\small{"CARMV" refers to CARMENES VIS and "CARMN" refers to CARMENES NIR.}} \\
\end{tabular}
\end{table*}

\section{RV data}
\label{appendixC}

\clearpage
\onecolumn
\begin{longtable}{crr}
\caption{RV data of CARMENES VIS.} \label{tab:RV_CARMV}\\
\hline\hline
Time & \multicolumn{1}{c}{RV} & \multicolumn{1}{c}{$\sigma$} \\

[BJD] & [m\,s$^{-1}$] & [m\,s$^{-1}$] \\
\hline
\endfirsthead
\caption{continued}\\
\hline\hline
Time & \multicolumn{1}{c}{RV} & \multicolumn{1}{c}{$\sigma$} \\

[BJD] & [m\,s$^{-1}$] & [m\,s$^{-1}$] \\
\hline
\endhead
\hline
\endfoot
\hline
\endlastfoot
2459111.6634 & 8.60 & 5.51\\
2459115.6509 & -28.62 & 6.86\\
2459119.6980 & -21.38 & 7.51\\
2459120.7016 & 13.87 & 3.40\\
2459121.6458 & -2.98 & 3.55\\
2459123.7132 & 16.28 & 4.20\\
2459127.7121 & 1.67 & 3.88\\
2459128.6559 & -16.55 & 4.06\\
2459131.6311 & -7.61 & 4.43\\
2459138.6096 & -16.80 & 3.77\\
2459139.6035 & 6.26 & 4.04\\
2459140.6105 & 3.02 & 4.34\\
2459141.5917 & -28.44 & 4.71\\
2459141.6521 & -27.37 & 4.93\\
2459146.5747 & 0.68 & 4.47\\
2459146.6579 & 5.12 & 3.81\\
2459147.5782 & -19.82 & 4.58\\
2459147.6578 & -20.68 & 3.53\\
2459149.6031 & 38.08 & 3.56\\
2459150.5866 & -2.78 & 3.94\\
2459151.6056 & 10.06 & 3.48\\
2459151.7019 & 12.49 & 3.61\\
2459152.6235 & 8.02 & 3.34\\
2459153.5519 & 12.61 & 3.83\\
2459154.5489 & -34.99 & 5.36\\
2459154.6326 & -24.76 & 4.21\\
2459156.5961 & -14.02 & 3.97\\
2459156.7021 & -7.86 & 3.78\\
2459161.5913 & 31.28 & 3.14\\
2459161.7395 & 45.88 & 3.84\\
2459162.5428 & 13.39 & 7.47\\
2459172.5011 & 2.39 & 4.08\\
2459172.5765 & 1.99 & 4.08\\
2459173.7197 & -13.32 & 3.81\\
2459174.5332 & 53.81 & 4.34\\
2459174.6128 & 57.16 & 3.78\\
2459174.7434 & 57.63 & 3.90\\
2459175.5304 & -6.59 & 3.22\\
2459175.6088 & -17.51 & 3.75\\
2459175.6951 & -19.95 & 2.86\\
2459177.5181 & 6.51 & 4.00\\
2459177.6235 & 5.57 & 3.34\\
2459177.6998 & 0.83 & 3.81\\
2459178.5291 & -19.34 & 4.81\\
2459178.6250 & -21.54 & 3.24\\
2459178.7445 & -16.35 & 4.34\\
2459183.4722 & 9.36 & 4.30\\
2459183.6414 & 7.59 & 3.87\\
2459183.7166 & 13.00 & 3.97\\
2459212.5915 & 12.29 & 5.78\\
2459474.6716 & -16.25 & 5.34\\
2459475.7023 & -1.60 & 3.69\\
2459483.6942 & 28.04 & 2.55\\
2459484.6693 & -8.58 & 4.35\\
2459485.7124 & 9.83 & 3.19\\
2459486.6918 & -6.66 & 3.59\\
2459488.7132 & 17.47 & 3.24\\
2459490.7157 & 24.52 & 3.52\\
2459492.7105 & 19.72 & 4.59\\
2459493.7101 & -14.29 & 3.78\\
2459494.6946 & 16.64 & 4.33\\
2459495.6974 & 37.17 & 2.91\\
2459502.7180 & 32.93 & 4.12\\
2459503.7056 & -8.82 & 4.02\\
2459506.6804 & -9.24 & 4.52\\
2459512.6666 & 4.69 & 4.31\\
2459513.5797 & -2.36 & 4.68\\
2459525.5196 & -12.60 & 4.15\\
2459525.5932 & -17.51 & 4.02\\
2459526.5250 & -13.02 & 3.39\\
2459526.5756 & -7.96 & 3.15\\
2459527.4875 & 5.95 & 3.94\\
2459527.5658 & -8.25 & 3.61\\
2459528.5737 & 23.32 & 3.95\\
2459529.5808 & -16.80 & 4.92\\
2459530.5470 & 13.79 & 3.31\\
2459534.6244 & -2.14 & 2.67\\
2459534.7331 & 8.91 & 3.15\\
2459535.5763 & -6.14 & 3.30\\
2459535.6534 & -3.59 & 3.23\\
2459548.4682 & -24.05 & 3.86\\
2459548.6102 & -27.83 & 3.26\\
2459549.4868 & 3.68 & 3.31\\
2459552.4621 & 3.44 & 3.68\\
2459553.4823 & -0.31 & 3.57\\
2459555.4819 & -27.26 & 4.22\\
2459559.6189 & -6.27 & 5.12\\
2459559.7245 & -5.76 & 4.93\\
2459560.5902 & 38.64 & 3.38\\
2459560.7285 & 40.16 & 3.36\\
2459561.5028 & -8.82 & 3.40\\
2459561.6415 & -12.89 & 3.12\\
2459562.6451 & 53.56 & 3.09\\
2459562.7227 & 61.24 & 3.42\\
2459563.5858 & 11.44 & 2.57\\
2459563.7244 & -4.37 & 3.45\\
2459564.6058 & -33.05 & 3.23\\
2459565.4960 & -6.66 & 4.47\\
2459565.5642 & -5.21 & 3.52\\
2459566.6345 & 2.42 & 4.13\\
2459568.5081 & 5.59 & 3.50\\
2459568.5696 & 6.18 & 3.39\\
2459573.5720 & 15.22 & 3.88\\
2459576.5090 & 1.36 & 3.76\\
2459576.5851 & 2.72 & 4.28\\
2459578.4434 & -1.15 & 4.00\\
2459578.5137 & -3.51 & 3.90\\
2459581.4516 & 11.43 & 3.87\\
2459581.5154 & 21.89 & 3.87\\
2459586.4345 & -32.14 & 4.71\\
2459586.5010 & -25.28 & 3.61\\
2459587.4953 & -27.94 & 3.68\\
2459588.3794 & 33.92 & 3.18\\
2459588.4898 & 35.63 & 3.18\\
2459589.3595 & -3.19 & 5.44\\
2459589.5088 & -4.73 & 5.71\\
2459590.4529 & -9.74 & 4.42\\
2459590.5298 & -14.51 & 3.71\\
2459595.5165 & 5.23 & 3.07\\
2459595.5878 & 5.19 & 3.06\\
2459596.5086 & -4.72 & 2.81\\
2459596.5889 & -8.88 & 3.39\\
2459597.5432 & -0.76 & 3.45\\
2459597.5928 & 5.10 & 3.27\\
2459603.5287 & -13.62 & 4.44\\
2459603.6096 & -12.79 & 4.43\\
2459609.4546 & -13.39 & 3.92\\
2459609.5174 & -17.93 & 3.68\\
2459613.5208 & 14.06 & 3.33\\
2459613.5736 & 14.52 & 3.43\\
2459614.3804 & 10.77 & 3.51\\
2459614.4739 & 12.68 & 3.93\\
2459614.5797 & 9.91 & 5.01\\
2459618.5332 & -9.24 & 3.20\\
2459618.5884 & -13.59 & 3.64\\
2459619.4324 & -5.76 & 3.50\\
2459619.5155 & 0.03 & 2.83\\
2459620.3800 & 15.97 & 3.65\\
2459620.4945 & 19.67 & 3.16\\
2459621.4036 & 15.89 & 4.39\\
2459621.5023 & 17.57 & 5.21\\
2459622.4144 & -8.86 & 3.89\\
2459623.4624 & -0.03 & 3.25\\
2459626.6579 & 4.74 & 4.98\\
2459629.4697 & -14.67 & 3.67\\
2459631.4810 & -15.92 & 2.97\\
2459632.4223 & -30.45 & 3.62\\
2459633.4158 & 19.53 & 3.40\\
2459634.4229 & 26.97 & 3.57\\
2459640.4633 & 38.56 & 3.45\\
\end{longtable}

\begin{center}
\begin{longtable}{crr}
\caption{RV data of CARMENES NIR.} \label{tab:RV_CARMN}\\
\hline\hline
Time & \multicolumn{1}{c}{RV} & \multicolumn{1}{c}{$\sigma$} \\

[BJD] & [m\,s$^{-1}$] & [m\,s$^{-1}$] \\
\hline
\endfirsthead
\caption{continued}\\
\hline\hline
Time & \multicolumn{1}{c}{RV} & \multicolumn{1}{c}{$\sigma$} \\

[BJD] & [m\,s$^{-1}$] & [m\,s$^{-1}$] \\
\hline
\endhead
\hline
\endfoot
\hline
\endlastfoot
2459111.6628 & 12.66 & 18.45\\
2459119.6980 & 16.34 & 31.05\\
2459120.7019 & 21.23 & 32.74\\
2459121.6460 & 6.83 & 8.27\\
2459123.7135 & 31.47 & 12.50\\
2459127.7119 & 9.76 & 12.64\\
2459128.6559 & 4.57 & 12.71\\
2459131.6312 & -12.58 & 14.19\\
2459138.6097 & 20.20 & 22.03\\
2459139.6036 & 18.65 & 16.78\\
2459140.6105 & 19.54 & 13.19\\
2459141.5918 & -9.47 & 13.40\\
2459146.5750 & 28.34 & 12.73\\
2459146.6579 & 41.66 & 12.65\\
2459147.5781 & -14.14 & 9.96\\
2459147.6582 & -27.41 & 14.15\\
2459149.6028 & 40.64 & 15.70\\
2459150.5866 & -5.00 & 14.93\\
2459151.7019 & 13.82 & 14.29\\
2459152.6234 & 1.73 & 12.34\\
2459154.5489 & -39.76 & 17.95\\
2459154.6326 & -26.81 & 10.10\\
2459156.5961 & -20.97 & 17.50\\
2459156.7019 & -15.50 & 13.63\\
2459161.5912 & 46.14 & 13.05\\
2459161.7396 & 76.60 & 11.51\\
2459172.5010 & -1.82 & 15.39\\
2459172.5764 & 5.40 & 12.99\\
2459173.7199 & -10.71 & 12.77\\
2459174.5336 & 58.07 & 14.34\\
2459174.6129 & 59.49 & 10.48\\
2459174.7432 & 61.78 & 13.51\\
2459175.5302 & -10.87 & 13.07\\
2459175.6086 & -18.64 & 13.21\\
2459175.6953 & -13.93 & 10.45\\
2459177.5183 & 12.54 & 12.80\\
2459177.6239 & -3.01 & 12.13\\
2459177.6998 & 10.34 & 14.44\\
2459178.5290 & -46.45 & 20.43\\
2459178.6253 & -27.73 & 12.14\\
2459178.7445 & -13.51 & 12.00\\
2459183.4723 & -3.94 & 13.19\\
2459183.6412 & 10.57 & 14.25\\
2459183.7163 & 3.54 & 14.15\\
2459212.5913 & -2.81 & 25.53\\
2459474.6718 & -23.88 & 19.14\\
2459475.7025 & -0.71 & 12.68\\
2459483.6944 & 43.14 & 11.09\\
2459484.6693 & -12.52 & 13.48\\
2459485.7124 & 10.17 & 11.40\\
2459486.6912 & 3.91 & 10.54\\
2459488.7125 & 22.76 & 10.61\\
2459490.7159 & 30.08 & 11.02\\
2459492.7106 & 31.44 & 14.60\\
2459493.7103 & -1.71 & 10.54\\
2459494.6947 & 29.69 & 11.77\\
2459495.6974 & 40.29 & 11.15\\
2459502.7173 & 66.33 & 11.73\\
2459503.7053 & 11.12 & 11.67\\
2459506.6805 & -3.03 & 11.83\\
2459512.6659 & 7.32 & 8.20\\
2459513.5803 & 25.69 & 29.59\\
2459525.5191 & 19.45 & 18.19\\
2459525.5929 & -3.86 & 17.78\\
2459526.5243 & 17.68 & 14.23\\
2459526.5755 & 9.32 & 13.67\\
2459527.4874 & 9.49 & 28.61\\
2459527.5658 & -0.99 & 11.92\\
2459528.5745 & 56.76 & 11.26\\
2459529.5817 & -2.39 & 13.86\\
2459530.5476 & 24.18 & 16.22\\
2459532.5377 & -28.52 & 17.21\\
2459533.5963 & 8.41 & 10.42\\
2459533.7247 & 7.55 & 9.94\\
2459534.6246 & 27.36 & 9.08\\
2459534.7327 & 30.08 & 10.77\\
2459535.5762 & -8.82 & 13.05\\
2459535.6535 & -25.34 & 10.23\\
2459548.4682 & -20.70 & 15.72\\
2459548.6101 & -37.82 & 8.21\\
2459549.4868 & 22.63 & 10.33\\
2459552.4618 & -6.97 & 13.68\\
2459553.4830 & -4.75 & 12.64\\
2459555.4818 & -37.78 & 15.58\\
2459559.6191 & 7.07 & 15.86\\
2459559.7245 & -6.18 & 14.88\\
2459560.5905 & 40.44 & 9.60\\
2459560.7284 & 54.29 & 11.42\\
2459561.5025 & -8.83 & 10.81\\
2459561.6412 & -18.45 & 10.28\\
2459562.6453 & 42.93 & 15.19\\
2459562.7226 & 56.19 & 12.46\\
2459563.5856 & 14.62 & 15.29\\
2459563.7242 & -3.80 & 10.28\\
2459564.6054 & -29.47 & 12.23\\
2459565.4956 & -9.12 & 10.63\\
2459565.5644 & -6.42 & 11.95\\
2459566.6352 & 4.40 & 15.81\\
2459568.5082 & 0.25 & 14.20\\
2459568.5701 & 32.11 & 14.06\\
2459573.5716 & 35.13 & 13.95\\
2459576.5091 & 1.37 & 12.51\\
2459576.5851 & -18.93 & 13.88\\
2459578.4433 & -17.49 & 21.50\\
2459578.5131 & 0.15 & 15.06\\
2459581.4517 & 9.03 & 13.52\\
2459581.5159 & 23.45 & 11.65\\
2459586.4347 & -19.66 & 16.41\\
2459586.5009 & -30.18 & 13.24\\
2459587.4952 & -25.20 & 12.57\\
2459588.3793 & 50.40 & 10.78\\
2459588.4895 & 57.39 & 9.97\\
2459589.3595 & -36.44 & 15.59\\
2459589.5083 & -19.74 & 37.74\\
2459590.4531 & -9.65 & 12.90\\
2459590.5301 & -27.67 & 13.55\\
2459595.5170 & 18.51 & 10.52\\
2459595.5879 & -14.22 & 10.34\\
2459596.5084 & -3.42 & 12.21\\
2459596.5889 & -6.54 & 9.33\\
2459597.5429 & -2.98 & 12.09\\
2459597.5923 & -1.52 & 11.72\\
2459603.5286 & -16.86 & 14.46\\
2459603.6096 & -17.61 & 14.50\\
2459609.5176 & -17.84 & 14.59\\
2459613.5212 & 5.95 & 16.43\\
2459613.5736 & -25.27 & 26.98\\
2459614.3807 & 12.84 & 13.34\\
2459614.4741 & -0.31 & 13.55\\
2459614.5801 & -2.96 & 17.36\\
2459618.5337 & -17.38 & 14.62\\
2459618.5883 & -35.69 & 14.11\\
2459619.4326 & -15.62 & 17.09\\
2459619.5154 & -25.00 & 14.12\\
2459620.3800 & 2.46 & 13.36\\
2459620.4943 & 9.64 & 13.04\\
2459621.4033 & -19.11 & 19.79\\
2459621.5023 & 12.34 & 27.95\\
2459622.4146 & -16.25 & 13.97\\
2459623.4621 & -1.84 & 14.61\\
2459629.4697 & -0.15 & 18.68\\
2459631.4803 & -25.48 & 15.02\\
2459632.4221 & -30.75 & 16.82\\
2459633.4157 & 5.66 & 13.65\\
2459634.4230 & 16.35 & 11.03\\
2459640.4634 & 28.52 & 16.83\\

\end{longtable}
\end{center}

\end{appendix}

\end{document}